\documentclass{article}
\usepackage{graphicx}

\usepackage{soul}
\usepackage{url}
\usepackage[hidelinks]{hyperref}
\usepackage[utf8]{inputenc}
\usepackage[small]{caption}
\usepackage{graphicx}
\usepackage{amsthm}
\usepackage{booktabs}
\usepackage{algorithm}
\usepackage{algorithmic}
\usepackage[cmex10]{amsmath}
\usepackage{scalerel}
\usepackage{amssymb}
\usepackage{mathtools}
\usepackage{color}
\usepackage{authblk}
\newtheorem{example}{Example}
\newtheorem{proposition}{Proposition}
\newtheorem{theorem}{Theorem}
\newtheorem{corollary}{Corollary}

\newtheorem{lemma}{Lemma}
\newtheorem{remark}{Remark}

\usepackage{tikz}
\usetikzlibrary{positioning,arrows,calc,shapes,babel,fit} 
\tikzset{
modal/.style={>=stealth',shorten >=1pt,shorten <=1pt,auto,node distance=1.5cm,
semithick},
world/.style={circle,draw,minimum size=0.5cm},
carg/.style={draw,minimum size=0.5cm},
point/.style={circle,draw,inner sep=0.5mm,fill=black},
reflexive above/.style={->,loop,looseness=7,in=120,out=60},
reflexive below/.style={->,loop,looseness=7,in=240,out=300},
reflexive left/.style={->,loop,looseness=7,in=150,out=210},
reflexive right/.style={->,loop,looseness=7,in=30,out=330}
}

\newcommand{\uni}{\mathcal{U}}
\newcommand{\aw}[1]{\mathtt{aw}_{#1}}
\newcommand{\acc}[1]{\mathtt{in}_{#1}}

\newcommand{\att}[2]{\mathtt{r}_{#1,#2}}

\newcommand{\argset}{A}
\newcommand{\attrel}{R}
\newcommand{\af}{(\argset,\attrel)}		

\newcommand{\semlist}{\{st, co, gr, pr, se, id, ea, na, stg\}}

\newcommand{\stableFml}{\mathsf{Stable}}

\newcommand{\stab}{\mathsf{st}}

\newcommand{\extOf}[1]{\mathsf{E}_{#1}} 

\newcommand{\assgntop}[1]{{+}#1}
\newcommand{\assgnbot}[1]{{-}#1}

\newcommand{\confree}{\mathsf{ConFree}}
\newcommand{\well}{\mathsf{Well}}
\newcommand{\mktrueone}{\mathsf{mkTrueOne}}
\newcommand{\mktruesome}{\mathsf{mkTrueSome}}
\newcommand{\mkfalseone}{\mathsf{mkFalseOne}}
\newcommand{\mkfalsesome}{\mathsf{mkFalseSome}}
\newcommand{\naive}{\mathsf{Naive}}
\newcommand{\pif}{\mathsf{if}}
\newcommand{\pthen}{\mathsf{then}}
\newcommand{\pelse}{\mathsf{else}}
\newcommand{\semistable}{\mathsf{SemiStable}}

\newcommand{\complete}{\mathsf{Complete}}
\newcommand{\pskip}{\mathsf{skip}}

\newcommand{\lbox}[1]{[#1]}
\newcommand{\ldia}[1]{\langle #1 \rangle}
\newcommand{\seq}{;}
\newcommand{\vary}{\mathsf{vary}}

\newcommand{\makecomp}{\mathsf{makeComp}}
\newcommand{\mkext}{\mathsf{makeExt}}
\newcommand{\control}{\mathsf{control}}

\newcommand\Seq{\scaleobj{1.75}{\seq}}

\newcommand{\infer}[2]      
    {{$\frac{\textstyle #1}{\textstyle #2}$} }

\newcommand{\limp}{\rightarrow}

\renewcommand{\phi}{\varphi}
\newcommand{\propset}     { \mathtt{Prp} }
\newcommand{\val}{\mathit{v}}

\newcommand{\dis}{\mathsf{dis}}
\newcommand{\awset}{\mathsf{AW}}
\newcommand{\attvarset}{\mathsf{ATT}}
\newcommand{\inset}{\mathsf{IN}}

\newcommand{\compfunc}{\mathsf{completions}}
\newcommand{\niaf}{\mathsf{IAF}}
\newcommand{\iaf}{(\fix,\unc)}
\newcommand{\unraveliaf}{(\argset^{\fix}\!,\attrel^{\fix}\!,\argset^{?}\!,\attrel^{?})}
\newcommand{\comp}{(\argset^{\ast},\attrel^{\ast})}

\newcommand{\nriaf}{\mathsf{rIAF}}
\newcommand{\symattrel}{\attrel^{\leftrightarrow}}

\newcommand{\riaf}{(\argset^{\fix},\attrel^{\fix},\argset^{?}\!, \attrel^{?}, \symattrel)}

\newcommand{\ncaf}{\mathsf{CAF}}
\newcommand{\caf}{(\fix,\unc,\con)}
\newcommand{\fix}{F}
\newcommand{\con}{C}
\newcommand{\unc}{U}
\newcommand{\cfg}{CFG}
\newcommand{\argsf}{\argset^{\fix}}
\newcommand{\attfix}{\attrel^{\fix}}

\newcommand{\nciaf}{\mathsf{cIAF}}
\newcommand{\ciaf}{(\argset,\varphi)}

\newcommand{\ncciaf}{\mathsf{CcIAF}}
\newcommand{\cciaf}{(\argset^{C}, \attrel^{C}, \argset^{S}, \varphi)}

\newcommand{\black}{\color{black}}

\begin{document}
\title{Qualitative Uncertainty and Dynamics of Argumentation through Dynamic Logic\footnote{This is a preliminary version of the paper ``Antonio Yuste-Ginel and Andreas Herzig, Qualitative uncertainty and dynamics of argumentation through dynamic logic, \emph{Journal of Logic and Computation}, 2023; exac098, \url{https://doi.org/10.1093/logcom/exac098}''. Please, consult the original publication and use the full reference for citation purposes.}}
\date{}
%

\author[1]{Antonio Yuste-Ginel} 

\author[1]{Andreas Herzig}

\affil[1]{IRIT, CNRS, Toulouse, France, herzig@irit.fr, antonioyusteginel@gmail.com}

\maketitle              
\begin{abstract}
Dynamics and uncertainty are essential features of real-life argumentation, and many recent studies have focused on integrating both aspects into Dung's well-known abstract Argumentation Frameworks (AFs). This paper proposes a combination of the two lines of research through a well-behaved logical tool: Dynamic Logic of Propositional Assignments (DL-PA). 
Our results show that the main reasoning tasks of virtually all existing formalisms qualitatively representing uncertainty about AFs are encodable in DL-PA. Moreover, the same tool is also useful for capturing dynamic structures, such as control argumentation frameworks, as well as for developing more refined forms of argumentative communication under uncertainty.

\end{abstract}

\sloppy

\section{Introduction}

Formal argumentation has been proved to be a successful approach to non-monotonic reasoning, among many other applications \cite{bench2007argumentation,atkinson2017towards,addedvalue}. Within the studies directed to provide a formal model for argument-based inference, abstract models of argumentation play a crucial role, as they answer a rather fundamental question: how should a rational agent choose among a conflicting set of arguments those that are better justified? The adjective \emph{abstract} stresses that these models disregard the nature and structure of arguments, in order to focus on the different semantics through which one could give a precise answer to the question above. The foremost abstract model of argumentation is the use of directed graphs, first proposed by Dung in \cite{dung1995acceptability} under the name of \emph{argumentation frameworks} (AFs), where nodes stand for arguments and arrows stand for attacks among arguments. \par 

While being an elegant and powerful tool, AFs have too limited modelling capabilities for many purposes. 
Consequently, many extensions of Dung's model were proposed in the literature, most prominently support relations \cite{cayrol2005acceptability}, recursive forms of attacks \cite{baroni2009encompassing}, and preferences between arguments \cite{amgoud2011new}. 
Two essential limitations of all these approaches are: (i)~their static character; and (ii)~the assumption that the formalized agent has perfect knowledge about the structure of the AF, that is, about the relevant arguments and attacks of the debate. \par 
 
Regarding (i), an AF can be understood as a snapshot of a debate, and this has been shown useful to provide mathematically precise counterparts of many interesting argumentative notions. However, a fundamental aspect of argumentation is its dynamic character, since arguments, conflicts among them, and participants' opinions typically change during the development of an argumentative dialogue. 
It is then unsurprising that the dynamics of formal argumentation systems 
has been the center of attention of a recent research avenue within formal argumentation, with an important focus on abstract models; we refer to \cite{DM18} and \cite{baumann2021enforcement} for recent surveys.

As to (ii), it turns out to be a significant shortcoming in adversarial contexts where one typically wants to model the information (i.e., the part of an AF) that an agent thinks her opponent entertains, and thus uncertainty arises naturally. 
This assumption of perfect knowledge has been relaxed through the study of extensions of AFs that account for different forms of uncertainty, be it probabilistic
or more qualitative; see \cite{hunter2021survey} and \cite{mailly2021yes} for recent surveys on the respective approaches. 
Among the second group of approaches, \emph{incomplete argumentation frameworks} (IAFs) \cite{baumeister2021acceptance,fazzing2020,baumeister2018credulous,baumeister2018verification,baumeister2018complexity} 
and \emph{control argumentation frameworks} (CAFs) \cite{dimopoulos2018control,niskanen2021controllability, cafsnegotiation} have recently received a lot of attention, resulting in a precise complexity map of the different associated reasoning tasks as well as some applications \cite{cafsnegotiation}.  

Concurrently, a considerable amount of work in formal argumentation has focused on building a suitable logical theory for reasoning about argumentation formalisms,
with a special focus on AFs and their dynamics; see \cite{besnard2020logical} for a recent survey.
The \emph{dynamic logic of propositional assignments} (DL-PA) \cite{balbiani2013dynamic} has been shown to be a useful tool for this enterprise \cite{doutre2014dynamic,doutre2017dynamic,doutre2019clar,clar2021}. 
DL-PA is a well-behaved variant of \emph{propositional dynamic logic} (PDL) \cite{HarelKozenTiuryn00}, where atomic programs are restricted to assignments of propositional variables to either Truth or Falsity. It is expressive enough to capture all standard argumentation semantics. When compared to encodings in propositional logic, DL-PA can capture semantics that incorporate minimality or maximality criteria more succinctly. 
Moreover, its advantages over equally succinct languages such as \emph{quantified Boolean formulas} have been highlighted \cite{doutre2019clar}.\par 

This work pushes further the logical encoding of abstract AFs in DL-PA
by pursuing three general aims:
(1)~to capture argumentation semantics that had not been captured before, some of them posing challenging encodings methods; 
(2)~to integrate qualitative uncertainty about AFs and dynamics of argumentation in DL-PA by reducing reasoning tasks of different extensions of argumentation frameworks to DL-PA model checking problems; 
and (3)~to show that the chosen logic is also a suitable tool for exploratory purposes, by developing new forms of modelling argumentative communication under uncertainty that are directly inspired by our encodings.

\par 
After providing the essential background on AFs and DL-PA (Section~\ref{sec:background}), Section~\ref{sec:semantics} provides polynomial encodings of a wide range of AF semantics in DL-PA. In Section~\ref{sec:uncertainty} we present several formalisms for qualitatively representing uncertainty about AFs, as well as the reduction of their main reasoning tasks to DL-PA model checking problems. In Section~\ref{sec:dynamics}, we discuss joint approaches to dynamics and qualitative uncertainty of AFs. 
Section~\ref{sec:final} ends the paper with some discussion and challenges for future work. 
Proofs and proof sketches can be found in the \nameref{sec:app}.

\section{Background}\label{sec:background}

Throughout the paper we assume a fixed, finite, non-empty set of arguments $\uni$ (the \emph{universe}). 
We moreover assume that $\uni$ is big enough to accommodate our examples. 
Sets of arguments (noted $\argset$, sometimes with a superscript) are supposed to be subsets of $\uni$; and all conflict relations (noted $\attrel$, sometimes with a superscript) are binary relations on $\uni$, 
i.e., $\attrel \subseteq \uni \times  \uni$. 
Given $\argset \subseteq \uni$ and $\attrel \subseteq \uni \times \uni$, we use $\attrel\upharpoonright_{ \argset}$ to abbreviate $\attrel \cap (\argset \times \argset)$ (the \textbf{restriction of $\attrel$ to $\argset$}). 

\subsection{Abstract Argumentation Frameworks (AFs) and their Semantics}\label{sec:background:argsema}

An \textbf{argumentation framework} (AF) is a directed graph $(\argset,\attrel)$ \cite{dung1995acceptability}, where $\argset$ stands for a set of arguments and $\attrel$ stands for a conflict-based relation among them (typically an attack relation).\footnote{
As $\argset\subseteq \uni$, we actually focus on \emph{finite} AFs, as most of the literature does. 
This is an inherent limitation of our approach: our encodings use quantification over $\uni$, which makes finiteness of $\uni$ necessary. Capturing some more general argumentation semantics has turned out to require powerful logical languages, such as the modal $\mu$-calculus for the grounded semantics \cite{grossi2010logic}.}
We note $\mathcal{AF}$ the set of all argumentation frameworks (over $\uni$).
Argumentation semantics are meant to capture the informal notion of reasonable positions in a debate. 
The literature contains a large number of such semantics. They are typically presented either in extension-based terms or in labelling-based terms. For most of the existing semantics, both approaches (extensions and labellings) were proved equivalent. Here, we opt for an extension-based presentation and restrict our attention to a limited number of semantics, but the interested reader is referred to \cite{baroni2018abstract} for an overview. 
 
Let us first define some useful concepts. Let $\af$ be an AF and let $E\subseteq \argset$. We define $E^{+}=\{x \in \argset \mid \exists y \in E: (y,x)\in \attrel\}$ (the \textbf{set of arguments attacked by $E$}), and $E^{\oplus}=E\cup E^{+}$ (the so-called \textbf{range of $E$}). A set of arguments $E\subseteq \argset$ is \textbf{conflict-free} iff $E\cap E^{+}=\emptyset$. Moreover, $E$ \textbf{defends} $a\in \argset$ iff for every $x \in \argset$: if $(x,a)\in \attrel$, then $x \in E^{+}$. Finally, $E\subseteq \argset$ is \textbf{admissible} iff it is (i) conflict-free and (ii) self-defended (it defends all its members).

In \cite{dung1995acceptability}, Dung introduced four different semantics. A set of arguments $E\subseteq \argset$ is said to be:

\begin{itemize}

\item a \textbf{stable extension} iff 
(i)~it is conflict-free, and 
(ii)$\argset \setminus E\subseteq E^{+}$ (`$E$ attacks every argument outside itself'); 

\item a \textbf{complete extension} iff 
(i)~it is conflict-free; and 
(ii)~it contains precisely the arguments of $\argset$ that it defends; 

\item a \textbf{grounded extension} iff it is a minimal (w.r.t.\ set inclusion) complete extension; 

\item a \textbf{preferred extension} iff it is a maximal (w.r.t.\ set inclusion) complete extension.
\end{itemize}
It is well known that the existence of complete, grounded and preferred extensions is guaranteed for any AF. However, this does not hold for stable semantics: there exist frameworks lacking stable extensions. Moreover, the grounded semantics is the only one from the above list {belonging to the so-called single-status approach}: each AF has exactly one grounded extension. This is an advantage when, for instance, modelling the beliefs of an agent as the output of her argument-evaluation processes. 
More precisely, if a semantics admits AFs with several extensions then these extensions are usually logically incompatible when one works with structured arguments, and there is no clear way to choose among them. 

Besides Dung's above four semantics we will take into account some others.
  
Semi-stable semantics was born to solve the problems caused by the absence of stable extensions under certain conditions. A set $E\subseteq \argset$ is a \textbf{semi-stable extension} of $\af$ iff $E$ is a complete extension with maximal (w.r.t.\ set inclusion) range among complete extensions. More formally, $E$ is a semi-stable extension iff

\begin{itemize}
\item[(i)] $E$ is a complete extension; and
\item[(ii)] there is no other complete extension $E'$ such that $E^{\oplus}\subset E'^{\oplus}$.
\end{itemize}

Contrarily to what happens with stable extensions, there is at least one semi-stable extension in every finite AF. Moreover, when the set of stable extensions is nonempty, stable and semi-stable extensions coincide \cite{caminada2012semi}. 

Although appealing because of its single-status approach, grounded semantics can be criticised as being too sceptical because it typically leaves many undecided arguments, i.e., arguments neither belonging to the grounded extension nor attacked by it. 
The idea of both the eager and the ideal semantics is to keep the advantage of returning a single extension while avoiding being overly sceptical. 

Formally, a set $E\subseteq A$ is an \textbf{ideal set} of $\af$ iff it is admissible and it is contained in every preferred extension. The \textbf{ideal extension} of $\af$ is its maximal (w.r.t.\ set inclusion) ideal set. 
 
Moreover, a set $E\subseteq A$ is an \textbf{eager set} iff it is admissible and it is contained in every semi-stable extension. The \textbf{eager extension} of $\af$ is its maximal (w.r.t.\ set inclusion) eager set.

All the above semantics satisfy the so-called \emph{admissibility principle}, meaning that all of their extensions are admissible sets. For some purposes, however, self-defence could be too strong a requirement, for instance when capturing human argument evaluation {\cite{guillaume2022plosone}}. Alternative semantics selecting specific conflict-free sets were defined under the denomination \emph{naivety-based} semantics (see e.g.\ \cite{cramer2019scf2}). 
The basis of all these semantics is the notion of naive extension. A \textbf{naive extension} of $(\argset,\attrel)$ is just a maximal (w.r.t.\ set inclusion) conflict-free set. {A more elaborated naivety-based semantics, strongly inspired in the notion of semi-stability, is stage semantics. Formally, a \textbf{stage extension} of $\af$ is a conflict-free set with maximal range among conflict-free sets.}

We abbreviate the name of each semantics by using the shorthands $\semlist$ in the obvious way. 
For every $\sigma\in \semlist$, we note $\sigma(\argset,\attrel)$ the set of all $\sigma$-extensions of $(\argset,\attrel)$. An argument $x\in \argset$
is said to be credulously (resp.\ sceptically) $\sigma$-\textbf{accepted} iff it belongs to at least one (resp.\ every) $\sigma$-extension.

As an example, for the AF $(\argset_0,\attrel_0)$ represented in the picture below we have 
$\stab(\argset_0,\attrel_0)=\mathsf{pr}(\argset_0,\attrel_0)=\mathsf{se}(\argset_0,\attrel_0)= \mathsf{stg}(\argset_0,\attrel_0)=\{\{b,e\},\{c,d\}\}$; $\mathsf{gr}(\argset_0,\attrel_0)=\mathsf{id}(\argset_0,\attrel_0)=\mathsf{ea}(\argset_0,\attrel_0)=\{\emptyset\}$; $\mathsf{co}(\argset_0,\attrel_0)=\{\emptyset,\{b,e\},\{c,d\}\}$ and $\mathsf{na}(\argset_0,\attrel_0)=\{\{a,c\},\{a,e\},\{b,e\},\{b,d\},\{c,d\}\}$.
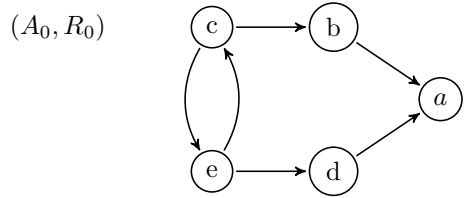
\begin{center}
\begin{tikzpicture}[modal,world/.append style=
{minimum size=0.5cm}]
\node[world] (a) [] {$a$};
\node (pos) [left=1 cm of a]{};
\node[world] (b) [above=0.5 cm of pos]{b};
\node[world] (c) [left=1 cm of b]{c};
\node (name) [left=1 cm of c]{$(\argset_0,\attrel_0)$};
\node[world] (d) [below=0.5 cm of pos]{d};
\node[world] (e) [left=1 cm of d]{e};
\draw[->] (b) edge (a);
\draw[->] (d) edge (a);
\draw[->] (c) edge (b);
\draw[->] (e) edge (d);
\draw[->] (c) edge[bend right] (e);
\draw[->] (e) edge[bend right] (c);
\end{tikzpicture}
\end{center}
Moreover, $(\argset_1,\attrel_1)$, depicted below and borrowed from \cite{caminada2012semi} illustrates the difference between stable and semi-stable semantics: the framework has no stable extension, while $\{c,a\}$ is a semi-stable extension (which is actually the only one). 

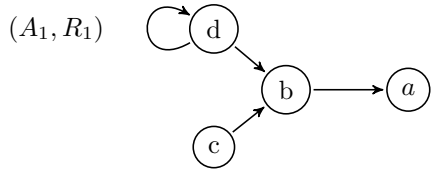
\begin{center}
\begin{tikzpicture}[modal,world/.append style=
{minimum size=0.5cm}]
\node[world] (a) [] {$a$};
\node[world] (b) [left=1 cm of a]{b};
\node (pos) [left=0.5 cm of b]{};
\node[world] (d) [above=0.35 cm of pos]{d};
\node (name) [left=1 cm of d]{$(\argset_1,\attrel_1)$};
\node[world] (c) [below=0.35 cm of pos] {c};

\draw[->] (b) edge (a);
\draw[->] (c) edge (b);
\draw[->] (d) edge (b);
\draw[->] (d) edge[reflexive left] (d);
\end{tikzpicture}
\end{center}

Finally, to see the difference between (semi-)stable and preferred semantics consider $(\argset_2,\attrel_2)$, borrowed from \cite{baroni2018abstract} and depicted below. The set $\{a\}$ is a preferred extension, but it is not a (semi-)stable one. Moreover, the example also illustrates the difference between ideal and eager semantics, as the eager extension is $\{b,d\}$, while the ideal one is empty. For more examples, the reader is referred to \cite{baroni2018abstract}, as well as to the graphic on-line solver \emph{ConArg} \cite{conargpaper}.
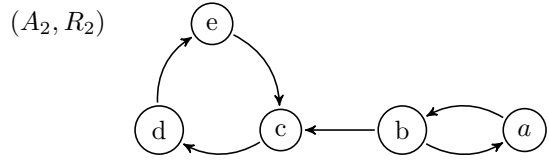
\begin{center}
\begin{tikzpicture}[modal,world/.append style=
{minimum size=0.5cm}]
\node[world] (a) [] {$a$};
\node[world] (b) [left=1 cm of a]{b};
\node[world] (c) [left=1 cm of b]{c};
\node[world] (d) [left=1 cm of c]{d};
\node (pos) [left=0.5 cm of c] {};
\node[world] (e) [above=1 cm of pos]{e};
\node (name) [left=1 cm of e]{$(\argset_2,\attrel_2)$};
\draw[->] (b) edge[bend right] (a);
\draw[<-] (b) edge[bend left] (a);

\draw[<-] (c) edge (b);
\draw[->] (d) edge[bend left] (e);
\draw[<-] (c) edge[bend right] (e);
\draw[->] (c) edge[bend left] (d);
\end{tikzpicture}
\end{center}

\subsection{Dynamic Logic of Propositional Assignments (DL-PA)}\label{sec:dlpa}

We use DL-PA as the general logical framework of this paper. 
The language of DL-PA is built from a {countably infinite} set of propositional variables $\propset = \{p_1, p_2, \ldots\}$. 
We suppose that $\propset$ contains several kinds of propositional variables capturing statuses of arguments and relations between them. 
First, to every set of arguments $\argset \subseteq \uni$ we associate the set of \textbf{awareness variables} 
$\awset_{\argset}=\{\aw x \mid x \in \argset\}$
and the set of \textbf{acceptance variables} 
$\inset_{\argset}=\{\acc x \mid x \in \argset\}$. 
Second, to every relation $\attrel \subseteq \uni \times \uni$ we associate the set of \textbf{attack variables}
$\attvarset_{\attrel}=\{\att x y \mid (x,y)\in \attrel\}$.
The set of propositional variables of our logic therefore contains 
\begin{align*}
\propset_{\uni} &= \awset_{\uni} \cup \inset_{\uni} 
\cup \attvarset_{\uni\times\uni} \text{.}
\end{align*}

As $\propset_{\uni}$ is finite, the countably infinite $\propset$ provides a reservoir of auxiliary variables that are going to help us to encode e.g.\ semi-stable and stage semantics.
Formulas and programs of DL-PA are defined by mutual recursion:
\begin{align*}
\varphi &::= p \mid \lnot \varphi \mid (\varphi \land \varphi)\mid [\pi]\varphi ,
\\
\pi &::= \assgntop{p} \mid \assgnbot{p} \mid \varphi?\mid (\pi;\pi) \mid (\pi \cup \pi) \mid \pi^{\smallsmile} ,
\end{align*}
where $p$ ranges over $\propset$.
The formula $[\pi]\varphi$ reads ``$\varphi$ is true after every possible execution of $\pi$''. 
The program $\assgntop{p}$ makes $p$ true and $\assgnbot{p}$ makes $p$ false. 
The program $\varphi?$ tests that $\varphi$ is true and fails when it is false. 
The program $\pi_1 ; \pi_2 $ is the sequential composition of $\pi_1$ and $\pi_2$; and 
$\pi_1 \cup \pi_2$ is their nondeterministic composition. 
Finally, $\pi^{\smallsmile} $ is the execution of $\pi$ `the other way round'; for example, the program $\assgntop p^{\smallsmile} $ undoes the assignment of $p$ to true: when $p$ is false then it fails, and when $p$ is true then it nondeterministically either does nothing or makes $p$ false. 

Here are some more examples. 
The formula $[\assgnbot p]\lnot p$ is going to be valid: there is only one way of executing $\assgnbot p$, and $p$ is false afterwards. 
In contrast, $[\assgntop p]\lnot p$ is going to be unsatisfiable. 
Moreover, $[\assgntop p]q$ is equivalent to $q$ for syntactically different $p$ and $q$. 
The formula $[\phi ? ] \psi$ says that $\psi$ is true after every possible execution of the test $\phi ? $. 
There is at most one such execution, namely when $\phi$ is true, and it does not change anything; when $\phi$ is false then the test fails and $[\phi ? ] \psi$ is vacuously true. 
Therefore $[\phi ? ] \psi$ has to be equivalent to $\lnot \phi \lor \psi$. 
The formula $[\pi_1 ; \pi_2]\phi$ is equivalent to $[\pi_1][\pi_2]\phi$ and $[\pi_1 \cup \pi_2]\phi$ is going to be equivalent to $[\pi_1]\phi \land [\pi_2]\phi$. 
Finally, the formulas $[\assgntop p \cup \assgnbot p]\lnot p$ and $[\assgnbot p^\smallsmile]p$ are both going to be unsatisfiable. The former is the case because there is a nondeterministic choice (namely that of $\assgntop p$) after which $p$ is true; the latter is the case because there is an execution of the nondeterministic $\assgnbot p^\smallsmile$ after which $p$ is still false. 

Here are some abbreviations of formulas and programs that are going to be useful in the rest of the paper. 
The program $\top?$ is abbreviated as $\pskip$: it always succeeds and does not change anything.
The program $(\varphi ? ; \alpha)\cup(\lnot \varphi ? ; \beta)$ abbreviates 
$\pif\, \varphi\, \pthen\, \alpha\, \pelse\, \beta$. 
A special case of the latter is when $\beta$ is $\pskip$, where we just write $\pif\, \phi\, \pthen\, \pi$.
(Observe that this is not the same as $\varphi ? ; \alpha$: when $\phi$ is false then the latter fails while the former succeeds and does nothing.) 
As to formulas, the missing Boolean connectives are defined as usual. 
Moreover, the formula $\ldia{\pi}\phi$ abbreviates $\lnot [\pi] \lnot \phi$. 
It therefore reads ``$\varphi$ is true after some possible execution of $\pi$''. 
In particular, $\ldia{\pi}\top$ has to be read ``$\pi$ is executable''.
For example, the formula $\phi \limp [\pi] \ldia{\pi^\smallsmile}\phi$ expresses that every successful execution of $\pi$ can be reversed; it is going to be valid. 
(Observe that the diamond cannot be replaced by a box, as illustrated by the invalid
$p \limp [\assgntop p] [\assgntop p^\smallsmile]p$.)

Our models are classical propositional valuations over $\propset$, i.e., they are subsets of $\propset$.
We use $\val,\val',\val''$ to denote valuations. 
Formulas $\varphi$ are interpreted in a way similar to dynamic logic, and programs $\pi$ are interpreted as binary relations on valuations. 
Just as the syntax, the semantics of DL-PA is defined by mutual recursion.
The interpretation of formulas is: 

\begin{center}\begin{tabular}{lcl}
$\val \models p$ &if& $p \in \val$,
\\
$\val \models [\pi]\varphi$ &if& $(\val,\val')\in ||\pi||$ implies $\val'\models \varphi$,
\end{tabular}\end{center}
and as usual for the Boolean connectives; 
and the interpretation of programs 
is:
\begin{align*}
||\assgntop{p}|| &=\{(\val,\val')\mid \val'=\val\cup\{p\}\} ,
\\ 
||\assgnbot{p}|| &=\{(\val,\val')\mid \val'=\val\setminus\{p\}\} ,
\\
||\varphi?|| &=\{(\val,\val)\mid \val\models \varphi\} ,
\\ 
||\pi; \pi'|| &=||\pi||\circ ||\pi'|| ,
\\
||\pi\cup \pi'|| &=||\pi||\cup ||\pi'|| ,
\\ 
||\pi^{\smallsmile}|| &=||\pi||^{-1} .
\end{align*}

The interpretation of $\assgntop p$ is the relation that makes $p$ true while not changing anything else; and similarly for $\assgnbot p$. 
That of the test $\phi ? $ relates every valuation where $\phi$ is true with itself; for example, 
$||\top ? || = \{(\val,\val) \mid \val \subseteq \propset\}$. 
Sequential composition $\pi_1 ; \pi_2$ is naturally interpreted as relation composition and nondeterministic composition $\pi_1 \cup \pi_2$ as set union of the two relations $||\pi_1||$ and $||\pi_2||$. 
The interpretation of the converse $\pi^\smallsmile$ is the inverse of the relation $||\pi||$ and relates a valuation $\val$ to all those valuations where $\pi$ is executable and may lead to $\val$. 
For example, $||\assgntop p^{\smallsmile}||=\{(\val',\val)\mid \val'=\val \cup \{p\}\}$.
\par
A formula $\varphi$ is DL-PA satisfiable if $\val \models \phi$ for some $\val$, and it is DL-PA valid if $\val \models \phi$ for every $\val$. 
For example, 
$\ldia{\assgntop p} \top$ and $\ldia{\assgnbot p} \top$ are both valid, while
$\ldia{\assgnbot p^\smallsmile} p$ and $\ldia{\assgnbot p^\smallsmile} \lnot p$ are satisfiable but not valid. 
It is known that satisfiability, validity, and model checking are all PSPACE complete decision problems \cite{BalbianiHST14}. 
 
Let us now introduce some DL-PA programs that will be useful later on. 
Let $\mathsf{P} = \{p_1,\ldots,p_n\} \subseteq \propset$ be a finite set of propositional variables. 
First of all, we define 
$$\Seq_{p\in \mathsf{P}} \pi_p = \pi_{p_1};\ldots;\pi_{p_n} . $$
{By convention, we assume that the abbreviation amounts to $\pskip$ when $\mathsf{P}=\emptyset$. We adopt the same convention for nondeterministic union, i.e., $\bigcup_{p \in \emptyset}\pi_p=\pskip$.} 
In principle the order of the elements of $\mathsf{P}$ matters, but each time we are going to use this notation we will make sure that the programs $\pi_{p_i}$ are such that this is not the case. 
This will in particular hold for the following abbreviations:
\begin{align*}
\mktrueone(\mathsf{P})&=\bigcup_{p\in \mathsf{P}}(\lnot p?; \assgntop p)=(\lnot p_1?;\assgntop {p_1})\cup \ldots \cup(\lnot p_n?;\assgntop{p_n}) \text{,} 
\\ 
\mkfalseone(\mathsf{P})&=\bigcup_{p\in \mathsf{P}}(p?; \assgnbot p)=(p_1?;\assgnbot {p_1})\cup \ldots \cup(p_n?;\assgnbot{p_n}) \text{,} 
\\ 
\mktruesome(\mathsf{P})&=\Seq_{p\in \mathsf{P}}(\assgntop p \cup \pskip)=(\assgntop{p_1} \cup \pskip); \ldots ;(\assgntop{p_n}\cup \pskip) \text{,}
\\
\mkfalsesome(\mathsf{P})&=\Seq_{p\in \mathsf{P}}(\assgnbot p \cup \pskip)=(\assgnbot{p_1} \cup \pskip); \ldots ;(\assgnbot{p_n}\cup \pskip) \text{,}
\\
\vary(\mathsf{P}) &=\Seq_{p\in \mathsf{P}}(\assgntop p \cup \assgnbot p) =\big(\assgntop{p_1}\cup \assgnbot{ p_1 }\big); \ldots ;\big(\assgntop{p_n} \cup \assgnbot{p_n}\big) .
\end{align*}
The program $\mktrueone(\mathsf{P})$ chooses an element of $\mathsf{P}$, checks that it is false and makes it true, while 
$\mktruesome(\mathsf{P})$ makes true some elements of $\mathsf{P}$ that were false before.
The programs $\mktrueone(\mathsf{P})$ and $\mkfalseone(\mathsf{P})$ are the converse of each other; same for 
$\mktruesome(\mathsf{P})$ and $\mkfalsesome(\mathsf{P})$. 
The sequential composition $\mktrueone(\mathsf{P}) ; \mktruesome(\mathsf{P})$ makes true at least one element of $\mathsf{P}$ that was false before (possibly more). 
The last program---i.e., $\vary(\mathsf{P})$---has the same interpretation as the sequential compositions 
$\mktruesome(\mathsf{P}) ; \mkfalsesome(\mathsf{P})$ and 
$\mkfalsesome(\mathsf{P}) ; \mktruesome(\mathsf{P})$.
\par 
Let us state formally the meaning of these programs:\footnote{The following proposition is a slight correction of \cite[Lemma 1]{doutre2019clar}.}
\begin{proposition}\label{prop:dlpaprg}
We have: 
\begin{align*}
||\mktrueone(\mathsf{P})|| &= \{(\val,\val') \mid \val' = \val \cup \{p\} \text{ for some } p \in \mathsf{P} {\setminus \val}\} ,
\\
||\mkfalseone(\mathsf{P})|| &= \{(\val,\val') \mid \val' = \val \setminus \{p\} \text{ for some } p \in \mathsf{P} {\cap \val}\} ,
\\ 
||\mktruesome(\mathsf{P})|| &= \{(\val,\val') \mid \val' = \val \cup \mathsf{P}' \text{ for some } \mathsf{P}' \subseteq \mathsf{P}\} ,
\\
||\mkfalsesome(\mathsf{P})|| &= \{(\val,\val') \mid \val' = \val \setminus \mathsf{P}' \text{ for some } \mathsf{P}' \subseteq \mathsf{P}\} ,
\\
||\vary(\mathsf{P})|| &= \{(\val,\val') \mid \val \setminus \val' \subseteq \mathsf{P} \text{ and } \val' \setminus \val \subseteq \mathsf{P} \} .
\end{align*}
\end{proposition}

\paragraph{From valuations to AFs and backward.} 
Thanks to our hypothesis that $\propset$ contains $\propset_{\uni}$, each valuation $\val\subseteq \propset$ represents the AF $(\argset_{\val},\attrel_{\val})$ defined by: 
\begin{align*}
\argset_\val &= \{x \in \uni \mid \aw x \in \val \} ,
\\
\attrel_\val &= \{(x,y)\in \uni \times \uni\mid \att x y \in \val\}\upharpoonright_{ \argset_{\val}}
\\
&= \{(x,y) \in \argset_\val \times \argset_\val \mid \att x y \in \val \} . 
\end{align*}
The other way round, each AF $(\argset,\attrel)$ is represented by the valuation
$$\val_{(\argset,\attrel)}=\{\aw x \mid x \in \argset\}\cup\{\att x y \mid (x,y)\in \attrel\}. $$
Note that the valuation $\val_{\af}$ is well defined for any set $\argset\subseteq \uni$ and relation $\attrel \subseteq \uni\times \uni$, even when $\af$ is not an AF.
(This is the case as soon as $\attrel$ contains pairs $(x,y) \in \uni\times\uni$ that are not in $\argset\times\argset$.)
Moreover, notice that if we start with a valuation $\val'$ then $\val_{(\argset_{\val'},\attrel_{\val'})}=\val'$ does not generally hold because a valuation can contain an attack variable $\att a b$ without containing $\aw a$ and $\aw b$. 
If we, however, start with an AF $(\argset',\attrel')$ then 
$(\argset_{\val_{(\argset',\attrel')}}, \attrel_{\val_{(\argset',\attrel')}})=(\argset',\attrel')$ 
is always the case. 
Finally, for each valuation $\val$ we define the \textbf{extension associated to $\val$} by:
$$\extOf\val = \{x \in \uni \mid \acc x \in \val\} . $$

\section{Argumentation Semantics in DL-PA}\label{sec:semantics}

We now show how to capture argumentation semantics in DL-PA. 
The starting point is to adopt the encoding of AFs in propositional logic as introduced in \cite{BesnardDoutre}. 
It consists in associating to each semantics $\sigma$ a formula $\varphi_{\sigma}$ such that $\val \models \varphi_{\sigma}$ if and only if $\extOf\val$ is a $\sigma$-extension of $(\argset_{\val},\attrel_{\val})$. 
This approach was pushed further in \cite{doutre2014dynamic,doutre2017dynamic,doutre2019clar,clar2021}, where it was proposed to go beyond the characterisation of extensions and exploit DL-PA programs to \emph{describe the computation of extensions}. 
The most basic way to do so is a `generate and test' approach: 
the generic program 
$$\mkext^{\sigma} = \vary(\inset_{\uni});\varphi_{\sigma}?$$ 
nondeterministically builds all possible $\sigma$-extensions by first varying the values of the acceptance variables and then checking that a $\sigma$-valuation has been obtained.
As worked out in \cite{doutre2019clar}, other, more efficient extension building algorithms can also be captured as DL-PA programs and can be proved to be equivalent to $\mkext^{\sigma}$. 

Due to our hypothesis of a background universe of arguments $\uni$ we need an encoding of argumentation semantics that takes awareness variables $\aw x$ into account. 
This was done by \cite{doutre2017dynamic} for stable semantics.\footnote{In \cite{doutre2017dynamic}, the term \emph{enablement} and the notation $\mathsf{En}_x$ are used instead of \emph{awareness} and $\aw x$.}  Here we extend the encoding to the rest of the semantics presented in Section~\ref{sec:background:argsema}. The correctness of all encodings is formally stated at the end of this section. 

We start by defining some formulas that allow us to capture the different semantics in a compact way.

\subsection{Useful DL-PA Formulas}

The following DL-PA formula expresses that the arguments identified by acceptance variables
are indeed arguments entertained by the formalised agent (arguments she is aware of): 
$$\well = \bigwedge_{x \in \uni}(\acc x \to \aw x)\text{.}$$
This abbreviation allows us to express conflict-freeness and admissibility:
\begin{align*}
\confree &= \well \land \bigwedge_{x\in \uni} \bigwedge_{y \in \uni} \lnot (\acc x \land \acc y \land \att x y)\text{,}
\\
\mathsf{Admissible} &= \confree \land \bigwedge_{x\in \uni} \Big(\acc x \to \bigwedge_{y \in \uni}\big((\aw y \land \att y x)\to \bigvee_{z \in \uni}(\acc z \land \att z y)  \big)\Big)\text{.}
\end{align*}

Our characterisation of semi-stable and stage extensions makes use of fresh copies $\acc x'$ of the variables $\acc x$, one per $x \in \uni$ (which are available because $\propset$ is countably infinite while $\uni$ is finite). For these auxiliary variables we define a program that copies the values of the $\inset_\uni$ variables:
\begin{align*}
\mathsf{copy}(\inset_{\uni}) &= \Seq_{x \in \uni} ((\acc{x} ? ; \assgntop{\acc{x}'}) \cup (\lnot \acc{x} ? ; \assgnbot{\acc x'}) ) .
\end{align*} 
Furthermore, the following two formulas characterise whether the range of the extension $\extOf\val$ represented by $\val$, in symbols $\extOf\val^{\oplus}$, is included in the range of the extension represented by the copies; and vice versa:
\begin{align*}
\mathsf{IncludedInCp} =\ & \bigwedge_{x \in \uni} 
\left[ \left(\acc{x} \lor \left(\aw{x} \land \bigvee_{y \in \uni}(\acc{y} \land \att y x) \right) \right) \right . \\& \qquad \limp
\left . \left(\acc{x}' \lor \left(\aw{x} \land \bigvee_{y \in \uni}(\acc {y}' \land \att y x) \right) \right) \right] ,
\\
\mathsf{IncludesCp} =\ & \bigwedge_{x \in \uni} \left[
\left(\acc{x}' \lor \left(\aw{x} \land \bigvee_{y \in \uni}(\acc{y}' \land \att y x) \right) \right) \right . \\& \qquad \limp
\left . \left(\acc{x} \lor \left(\aw{x} \land \bigvee_{y \in \uni}(\acc {y} \land \att y x) \right) \right) \right] .
\end{align*} 

Finally, to capture ideal and eager semantics we need to ensure that the entertained set is admissible and belongs to every preferred extension (for the case of ideal semantics), or to every semi-stable extension (for the case of eager semantics).
This can be done in a compact way by means of the extension-building programs $\mkext^{\sigma}$:

\begin{align*}
\mathsf{IdealSet} &= \mathsf{Admissible}\land \bigwedge_{x \in \uni}(\acc x \to [\mkext^{pr}] \acc x) ,
\\
\mathsf{EagerSet} &= \mathsf{Admissible}\land \bigwedge_{x \in \uni}(\acc x \to [\mkext^{se}] \acc x) .
\end{align*}

\subsection{Encoding the Semantics of Section~\ref{sec:background:argsema} in DL-PA}

Table~\ref{table:encodings} lists all the encodings.
That of stable and complete semantics slightly simplifies that of \cite{doutre2017dynamic,clar2021}.
Our encoding of grounded, complete, and preferred semantics straightforwardly adapts those of \cite{doutre2019clar} for computing minimality and maximality criteria.
The first four encodings are essentially a combination of those developed in \cite{doutre2017dynamic} and \cite{doutre2019clar}, with some slight improvements and adaptations. 
Among the semantics that have not been captured in DL-PA before, our encoding of naive semantics simplifies the program for checking set maximality w.r.t.\ other semantics such as preferred semantics because no superset of a set containing conflicts can be conflict-free. 
\begin{table}
\begin{align*}
\stableFml =\ &  \well \land \bigwedge_{x\in \uni} \Big( \aw x\to\big(\acc x \leftrightarrow \lnot \bigvee_{y \in \uni}(\acc y \land \att y x \big)\Big),
\\
\mathsf{Complete} =\ & \confree \land \bigwedge_{x\in \uni} \Big(\acc x \leftrightarrow \bigwedge_{y \in \uni}\big((\aw y \land \att y x)\to \bigvee_{z \in \uni}(\acc z \land \att z y)  \big)\Big),
\\
\mathsf{Grounded} =\ & \mathsf{Complete} \land [\mkfalseone(\inset_{\uni});\mkfalsesome(\inset_{\uni})] \lnot \mathsf{Complete} ,
\\
\mathsf{Preferred} =\ & \mathsf{Admissible} \land [\mktrueone(\inset_{\uni});\mktruesome(\inset_{\uni})] \lnot \mathsf{Admissible} ,
\\
\naive =\ & \confree \land [\mktrueone(\inset_{\uni})]\lnot \confree ,
\\
\semistable =\ & \complete \land 
\lbox{ \mathsf{copy}(\inset_{\uni}) ; \mkext^{co} } \left(
 \mathsf{IncludesCp} \limp \mathsf{IncludedInCp}  \right) ,
\\
\mathsf{Stage} =\ & \confree \ \land \\& \lbox{ \mathsf{copy}(\inset_{\uni}) ; \vary (\inset_{\uni});\confree? } \left(
 \mathsf{IncludesCp} \limp \mathsf{IncludedInCp}  \right) ,
\\
\mathsf{Ideal} =\ & \mathsf{IdealSet}\land [\mktrueone(\inset_{\uni});\mktruesome(\inset_{\uni})] \lnot \mathsf{IdealSet} ,
\\
\mathsf{Eager} =\ & \mathsf{EagerSet}\land [\mktrueone(\inset_{\uni});\mktruesome(\inset_{\uni})] \lnot \mathsf{EagerSet} .
\end{align*}
\caption{Encoding the Semantics of Section~\ref{sec:background:argsema} by DL-PA formulas}\label{table:encodings}
\end{table}

\begin{theorem}\label{thrm:encodings} Let $\sigma\in \semlist$. Let $\val\subseteq \propset$. Let $\af$ be an AF. Then:
\begin{itemize}
\item 
$\val\models \varphi_{\sigma}$ iff $\extOf \val \in \sigma(\argset_v,\attrel_v)$;
\item 
$\sigma(\argset,\attrel)=\big\{\extOf \val \mid (\val_{\af},v)\in ||\mkext^{\sigma}||\big\}$.
\end{itemize}
\end{theorem}
The proof can be found in the \nameref{sec:app}, just as the proofs or proof sketches of all other results.

\section{Qualitative Uncertainty in Abstract Argumentation through DL-PA}\label{sec:uncertainty} 

In this section, we review existing formalisms for representing uncertainty about AFs. We restrict our attention to \emph{qualitative} forms of uncertainty, that is, representations neither using probabilities nor any other kind of numeric device. 
In particular, we cover: \emph{incomplete argumentation frameworks} \cite{baumeister2021acceptance}, their \emph{enriched version} \cite{mailly2020note}, \emph{constrained incomplete argumentation frameworks} \cite{clar2021,maillyciafs}, and \emph{incomplete argumentation frameworks with dependencies} \cite{fazzingaijcai21,fazzingakr21}. 
The main motivation for the study of these formalisms is that there are several sources of uncertainty in real-life argumentation. For instance, arguments can be so complex that the reasoning agent is not sure whether they are to be taken into account or whether they attack other arguments. Perhaps more frequently, uncertainty appears in argumentation when an agent reasons about \emph{her opponent's argumentative situation}. Due to the lack of total knowledge about her adversary, the agent might doubt whether the latter entertains some of the arguments or sees some of the attacks. And this is in turn crucial for choosing the right arguments to convince her opponent. We keep this latter intuition in mind as a guideline for the rest of the paper.

All the formalisms of the present section share the idea of representing uncertainty through the notion of \emph{completion}. A completion is a hypothetical removal of uncertainty, such that the formalised agent reasons under the assumption that her opponent's AF is such-and-such. In epistemic logic terms, this amounts to the notion of \emph{possible world}, as mentioned in \cite{baumeister2018credulous,baumeister2018complexity}, and studied in detail in \cite{proietti2021,kr}. For a more elaborated comparison among the formalisms presented in this section and epistemic logic, the interested reader is referred to Section \ref{sec:final}.

After introducing each formalism we explain how the main associated reasoning tasks can be reduced to DL-PA model checking problems. We conclude by providing a comparison of the different approaches.

\subsection{Incomplete AFs}\label{sec:iafs}

An \textbf{incomplete AF} \cite{baumeister2021acceptance} (IAF), is a pair $\niaf=(\fix,\unc)$, where $\fix=(\argset^{\fix},\attrel^{\fix})$ is called the \emph{fixed part}, $\unc=(\argset^{?},\attrel^{?})$ is called the \emph{uncertain part}, $\attrel,\attrel^{?}\subseteq (\argset^{\fix}\cup \argset^{?})\times(\argset^{\fix}\cup \argset^{?})$, $\argset^{\fix}\cap\argset^{?}=\emptyset$ and $\attrel^{\fix}\cap\attrel^{?}=\emptyset$. 
Hence an IAF is basically an AF where arguments and attacks have been split into two disjoint sets. We sometimes omit internal parentheses when talking about IAFs, that is, we write $\unraveliaf$ instead of $((\argset^{\fix},\attrel^{\fix}),(\argset^{?},\attrel^{?}))$.
Note that, by definition, there can be fixed attacks among uncertain arguments (sometimes called \emph{conditionally definite attacks} \cite{baumeister2018credulous}). We can intuitively think about these as attacks the agent thinks her opponent entertains whenever she thinks that her opponent is aware of the involved arguments. \par

A \textbf{completion} of an $\niaf=\unraveliaf$ is any AF $\comp$ such that:
\begin{itemize}
\item $\argset^{\fix}\subseteq \argset^{\ast}\subseteq \argset^{\fix}\cup \argset^{?}$; and
\item $\attrel^{\fix}\upharpoonright_{ \argset^{\ast}}\subseteq \attrel^{\ast}\subseteq (\attrel^{\fix}\cup \attrel^{?})\upharpoonright_{ \argset^{\ast}}$.
\end{itemize}
Given an IAF $\niaf$, we note $\compfunc(\niaf)$ the set of all its completions. 

A standard AF $(\argset,\attrel)$ can be identified with the IAF $(\argset,\attrel,\emptyset,\emptyset)$, which is the unique completion of itself. 
Two subclasses of IAFs are well-studied in the literature, namely \textbf{attack-incomplete AFs} (att-IAFs, for short),\footnote{This subclass was previously studied under the name of \emph{partial AFs} \cite{cayrol2007partial,coste2007merging}.}  
which are IAFs with empty $\argset^{?}$; and \textbf{argument-incomplete} AFs (arg-IAFs, for short), which are IAFs with empty $\attrel^{?}$.

\begin{example}\label{ex:iaf} Let us consider $\niaf_0=(\argset^{\fix}_0,\attrel^{\fix}_{0},\argset^{?}_0,\attrel^{?}_0)$, where $\argset^{\fix}_0=\{a,b,d\}$, $\attrel^{\fix}_{0}=\{(b,a),(d,a),(c,b),(e,d),(c,e),(e,c), (f,e)\}$, $\argset^{?}_{0}=\{c,e,f\}$ and $\attrel^{?}_{0}=\{(f,c)\}$, graphically represented below. The set of completions of $\niaf_0$ is the one depicted in Table~\ref{tab:comp} except for the cells \textbf{B2}, \textbf{C2}, \textbf{B4}, \textbf{C4}, \textbf{B5} and \textbf{C5}. 

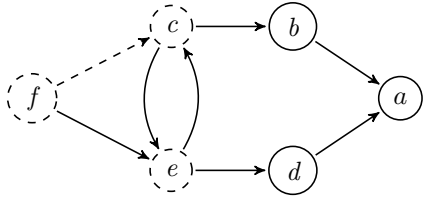
\begin{center}
\begin{tikzpicture}[modal,world/.append style=
{minimum size=0.5cm}]
\node[world] (a) [] {$a$};
\node (pos) [left=1 cm of a]{};
\node[world] (b) [above=0.5 cm of pos]{b};
\node[world,dashed] (c) [left=1 cm of b]{c};
\node[world] (d) [below=0.5 cm of pos]{d};
\node[world,dashed] (e) [left=1 cm of d]{e};
\node[world,dashed] (f) [left=3 cm of pos]{f};
\draw[->] (b) edge (a);
\draw[->] (d) edge (a);
\draw[->] (c) edge (b);
\draw[->] (e) edge (d);
\draw[->] (c) edge[bend right] (e);
\draw[->] (e) edge[bend right] (c);
\draw[->] (f) edge (e);
\draw[->] (f) edge[dashed] (c);
\end{tikzpicture}
\end{center}
\end{example}


\begin{table}
\footnotesize
\begin{tabular}{c| c | c | c|}
& \textbf{A} & \textbf{B} & \textbf{C} \\
\hline
& & &  \\
\textbf{1}& \begin{tikzpicture}[modal,world/.append style=
{minimum size=0.5cm}]

\node[world] (a) [] {$a$};
\node (pos) [left=0.5 cm of a]{};
\node[world] (b) [above=0.25 cm of pos]{b};
\node (c) [left=0.5 cm of b]{};
\node[world] (d) [below=0.25 cm of pos]{d};
\node (e) [left=0.5 cm of d]{};
\node (f) [left=1.5 cm of pos]{};
\draw[->] (b) edge (a);
\draw[->] (d) edge (a);

\end{tikzpicture}
&  \begin{tikzpicture}[modal,world/.append style=
{minimum size=0.5cm}]
\node[world] (a) [] {$a$};
\node (pos) [left=0.5 cm of a]{};
\node[world] (b) [above=0.25 cm of pos]{b};
\node[world] (c) [left=0.5 cm of b]{c};
\node[world] (d) [below=0.25 cm of pos]{d};
\node (e) [left=0.5 cm of d]{};
\node (f) [left=1.5 cm of pos]{};
\draw[->] (b) edge (a);
\draw[->] (d) edge (a);
\draw[->] (c) edge (b);
\end{tikzpicture} & 

\begin{tikzpicture}[modal,world/.append style=
{minimum size=0.5cm}]
\node[world] (a) [] {$a$};
\node (pos) [left=0.5 cm of a]{};
\node[world] (b) [above=0.25 cm of pos]{b};
\node (c) [left=0.5 cm of b]{};
\node[world] (d) [below=0.25 cm of pos]{d};
\node[world] (e) [left=0.5 cm of d]{e};
\node (f) [left=1.5 cm of pos]{};
\draw[->] (b) edge (a);
\draw[->] (d) edge (a);
\draw[->] (e) edge (d);
\end{tikzpicture}
 \\ 
 \hline
 & & & \\
 \textbf{2} & 
  \begin{tikzpicture}[modal,world/.append style=
{minimum size=0.5cm}]
\node[world] (a) [] {$a$};
\node (pos) [left=0.5 cm of a]{};
\node[world] (b) [above=0.25 cm of pos]{b};
\node[world] (c) [left=0.5 cm of b]{c};
\node[world] (d) [below=0.25 cm of pos]{d};
\node[world] (e) [left=0.5 cm of d]{e};
\node (f) [left=1.5 cm of pos]{};
\draw[->] (b) edge (a);
\draw[->] (d) edge (a);
\draw[->] (c) edge (b);
\draw[->] (e) edge (d);
\draw[->] (c) edge[bend right] (e);
\draw[->] (e) edge[bend right] (c);
\end{tikzpicture}
& 
 \begin{tikzpicture}[modal,world/.append style=
{minimum size=0.5cm}]
\node[world] (a) [] {$a$};
\node (pos) [left=0.5 cm of a]{};
\node[world] (b) [above=0.25 cm of pos]{b};
\node[world] (c) [left=0.5 cm of b]{c};
\node[world] (d) [below=0.25 cm of pos]{d};
\node[world] (e) [left=0.5 cm of d]{e};
\node (f) [left=1.5 cm of pos]{};
\draw[->] (b) edge (a);
\draw[->] (d) edge (a);
\draw[->] (c) edge (b);
\draw[->] (e) edge (d);
\draw[->] (c) edge[bend right] (e);

\end{tikzpicture}
&
 \begin{tikzpicture}[modal,world/.append style=
{minimum size=0.5cm}]
\node[world] (a) [] {$a$};
\node (pos) [left=0.5 cm of a]{};
\node[world] (b) [above=0.25 cm of pos]{b};
\node[world] (c) [left=0.5 cm of b]{c};
\node[world] (d) [below=0.25 cm of pos]{d};
\node[world] (e) [left=0.5 cm of d]{e};
\node (f) [left=1.5 cm of pos]{};
\draw[->] (b) edge (a);
\draw[->] (d) edge (a);
\draw[->] (c) edge (b);
\draw[->] (e) edge (d);

\draw[->] (e) edge[bend right] (c);
\end{tikzpicture} \\
\hline
& & & \\

 \textbf{3} & 

\begin{tikzpicture}[modal,world/.append style=
{minimum size=0.5cm}]
\node[world] (a) [] {$a$};
\node (pos) [left=0.5 cm of a]{};
\node[world] (b) [above=0.25 cm of pos]{b};
\node (c) [left=0.5 cm of b]{};
\node[world] (d) [below=0.25 cm of pos]{d};
\node (e) [left=0.5 cm of d]{};
\node[world] (f) [left=1.5 cm of pos]{f};
\draw[->] (b) edge (a);
\draw[->] (d) edge (a);

\end{tikzpicture} 
\quad
&
\quad
\begin{tikzpicture}[modal,world/.append style=
{minimum size=0.5cm}]
\node[world] (a) [] {$a$};
\node (pos) [left=0.5 cm of a]{};
\node[world] (b) [above=0.25 cm of pos]{b};
\node[world] (c) [left=0.5 cm of b]{c};
\node[world] (d) [below=0.25 cm of pos]{d};
\node (e) [left=0.5 cm of d]{};
\node[world] (f) [left=1.5 cm of pos]{f};
\draw[->] (b) edge (a);
\draw[->] (d) edge (a);
\draw[->] (c) edge (b);
\end{tikzpicture} 
\quad
& 
\quad
\begin{tikzpicture}[modal,world/.append style=
{minimum size=0.5cm}]
\node[world] (a) [] {$a$};
\node (pos) [left=0.5 cm of a]{};
\node[world] (b) [above=0.25 cm of pos]{b};
\node[world] (c) [left=0.5 cm of b]{c};
\node[world] (d) [below=0.25 cm of pos]{d};
\node (e) [left=0.5 cm of d]{};
\node[world] (f) [left=1.5 cm of pos]{f};
\draw[->] (b) edge (a);
\draw[->] (d) edge (a);
\draw[->] (c) edge (b);
\draw[->] (f) edge (c);
\end{tikzpicture} 
\\
\hline
& & & \\
 \textbf{4} & 
\begin{tikzpicture}[modal,world/.append style=
{minimum size=0.5cm}]
\node[world] (a) [] {$a$};
\node (pos) [left=0.5 cm of a]{};
\node[world] (b) [above=0.25 cm of pos]{b};
\node[world] (c) [left=0.5 cm of b]{c};
\node[world] (d) [below=0.25 cm of pos]{d};
\node[world] (e) [left=0.5 cm of d]{e};
\node[world] (f) [left=1.5 cm of pos]{f};
\draw[->] (b) edge (a);
\draw[->] (d) edge (a);
\draw[->] (c) edge (b);
\draw[->] (e) edge (d);
\draw[->] (c) edge[bend right] (e);
\draw[->] (e) edge[bend right] (c);
\draw[->] (f) edge (e);
\end{tikzpicture}
& 
 \begin{tikzpicture}[modal,world/.append style=
{minimum size=0.5cm}]
\node[world] (a) [] {$a$};
\node (pos) [left=0.5 cm of a]{};
\node[world] (b) [above=0.25 cm of pos]{b};
\node[world] (c) [left=0.5 cm of b]{c};
\node[world] (d) [below=0.25 cm of pos]{d};
\node[world] (e) [left=0.5 cm of d]{e};
\node[world] (f) [left=1.5 cm of pos]{f};
\draw[->] (b) edge (a);
\draw[->] (d) edge (a);
\draw[->] (c) edge (b);
\draw[->] (e) edge (d);
\draw[->] (c) edge[bend right] (e);
\draw[->] (f) edge (e);
\end{tikzpicture}
&
 \begin{tikzpicture}[modal,world/.append style=
{minimum size=0.5cm}]
\node[world] (a) [] {$a$};
\node (pos) [left=0.5 cm of a]{};
\node[world] (b) [above=0.25 cm of pos]{b};
\node[world] (c) [left=0.5 cm of b]{c};
\node[world] (d) [below=0.25 cm of pos]{d};
\node[world] (e) [left=0.5 cm of d]{e};
\node[world] (f) [left=1.5 cm of pos]{f};
\draw[->] (b) edge (a);
\draw[->] (d) edge (a);
\draw[->] (c) edge (b);
\draw[->] (e) edge (d);
\draw[->] (e) edge[bend right] (c);
\draw[->] (f) edge (e);
\end{tikzpicture} \\
\hline & & & \\
 \textbf{5} & 
\begin{tikzpicture}[modal,world/.append style=
{minimum size=0.5cm}]
\node[world] (a) [] {$a$};
\node (pos) [left=0.5 cm of a]{};
\node[world] (b) [above=0.25 cm of pos]{b};
\node[world] (c) [left=0.5 cm of b]{c};
\node[world] (d) [below=0.25 cm of pos]{d};
\node[world] (e) [left=0.5 cm of d]{e};
\node[world] (f) [left=1.5 cm of pos]{f};
\draw[->] (b) edge (a);
\draw[->] (d) edge (a);
\draw[->] (c) edge (b);
\draw[->] (e) edge (d);
\draw[->] (c) edge[bend right] (e);
\draw[->] (e) edge[bend right] (c);
\draw[->] (f) edge (e);
\draw[->] (f) edge (c);

\end{tikzpicture}

&
\begin{tikzpicture}[modal,world/.append style=
{minimum size=0.5cm}]
\node[world] (a) [] {$a$};
\node (pos) [left=0.5 cm of a]{};
\node[world] (b) [above=0.25 cm of pos]{b};
\node[world] (c) [left=0.5 cm of b]{c};
\node[world] (d) [below=0.25 cm of pos]{d};
\node[world] (e) [left=0.5 cm of d]{e};
\node[world] (f) [left=1.5 cm of pos]{f};
\draw[->] (b) edge (a);
\draw[->] (d) edge (a);
\draw[->] (c) edge (b);
\draw[->] (e) edge (d);
\draw[->] (c) edge[bend right] (e);

\draw[->] (f) edge (e);
\draw[->] (f) edge (c);

\end{tikzpicture}
& \begin{tikzpicture}[modal,world/.append style=
{minimum size=0.5cm}]
\node[world] (a) [] {$a$};
\node (pos) [left=0.5 cm of a]{};
\node[world] (b) [above=0.25 cm of pos]{b};
\node[world] (c) [left=0.5 cm of b]{c};
\node[world] (d) [below=0.25 cm of pos]{d};
\node[world] (e) [left=0.5 cm of d]{e};
\node[world] (f) [left=1.5 cm of pos]{f};
\draw[->] (b) edge (a);
\draw[->] (d) edge (a);
\draw[->] (c) edge (b);
\draw[->] (e) edge (d);

\draw[->] (e) edge[bend right] (c);
\draw[->] (f) edge (e);
\draw[->] (f) edge (c);
\end{tikzpicture}
\\
\hline
& & & \\
\textbf{6}
&
\begin{tikzpicture}[modal,world/.append style=
{minimum size=0.5cm}]
\node[world] (a) [] {$a$};
\node (pos) [left=0.5 cm of a]{};
\node[world] (b) [above=0.25 cm of pos]{b};
\node (c) [left=0.5 cm of b]{};
\node[world] (d) [below=0.25 cm of pos]{d};
\node[world] (e) [left=0.5 cm of d]{e};
\node[world] (f) [left=1.5 cm of pos]{f};
\draw[->] (b) edge (a);
\draw[->] (d) edge (a);
\draw[->] (f) edge (e);
\end{tikzpicture}
 & & \\
 & & & \\
\hline
\end{tabular}
\bigskip 
\caption{Completions of $\ncaf_0$. The column [\textbf{1}, \textbf{2},..., \textbf{6}] and the row [\textbf{A}, \textbf{B}, \textbf{C}] are just included for numbering purposes. 
(Empty cells do \emph{not} represent the empty completion $(\emptyset,\emptyset)$.)}
\label{tab:comp}
\end{table}
\normalsize

Classic reasoning tasks such as extension enumeration or argument acceptance have been generalized from AFs to IAFs. 
We here focus on acceptance queries such as the following: 

\begin{center}
\begin{tabular}{|l|}
\hline
$\sigma$-Necessary-Credulous-Acceptance ($\sigma$-NCA)\\
\hline

\textbf{Given:}  An IAF
 $\niaf=\unraveliaf$ 
  and an argument $a\in \argsf$. \\

\textbf{Question:} Is it true that for every 
$(\argset^{\ast},\attrel^{\ast}) \in \mathsf{completions}(\niaf)$ \\ there is an $E\in \sigma(\argset^{\ast},\attrel^{\ast})$ such that $a \in E$? \\
\hline
\end{tabular}
\end{center}

\noindent We can switch quantifiers in the definition above in order to obtain different variants of the problem, resulting in possible and sceptical variants. Note that the only difference between these reasoning tasks and standard acceptance problems in AFs is an added quantification layer, namely quantification over completions. 
 
\smallskip
Our aim now is to reduce these acceptance problems to DL-PA model checking problems. As we already have programs for building the extensions of AFs, the fundamental step in this reduction consists in designing a DL-PA program, $\makecomp^{\niaf}$, that computes all the completions of $\niaf$.

First, the \textbf{valuation associated to $\niaf$} is determined by its fixed part:
\begin{align*}
v_{\niaf}&=v_{(\argset^{\fix},\attrel^{\fix})} 
\\&= \awset_{\argset^{\fix}} \cup \attvarset_{\attrel^{\fix}}
\\&= \{\aw x \mid x \in \argsf\}\cup\{\att x y \mid (x,y) \in \attfix\} .
\end{align*}

Note that $(\argset_{\val_{\niaf}},\attrel_{\val_{\niaf}})$ is already a completion of $\niaf$: 
it is the smallest one, where only fixed arguments and fixed attacks between them are considered. 
In order to compute all the completions of $\niaf$ we make true subsets of propositional variables representing arguments in $\argset^{?}$ and attacks in $\attrel^{?}$: 
\begin{align*}
\makecomp^{\niaf}&=\mktruesome(\awset_{\argset^{?}}) \seq \mktruesome(\attvarset_{\attrel^{?}})\text{.}
\end{align*}
The next proposition shows that our original target is reached.
\begin{proposition}\label{prop:iafenco} Let $\niaf=\unraveliaf$. Then: 
\begin{itemize}
\item If $(\val_{\niaf},\val)\in ||\makecomp^{\niaf}||$, then $(\argset_{\val},\attrel_{\val})\in \mathsf{completions}(\niaf)$.
\item If $(\argset^{\ast},\attrel^{\ast})\in \mathsf{completions}(\niaf)$, then 
$(\val_{\niaf},\val_{(\argset^{\ast},\attrel^{\ast})})\in ||\makecomp^{\niaf}||$.
\end{itemize} \end{proposition}

Using this result together with the general technique to compute extensions provided in Section~\ref{sec:semantics}, we can reduce reasoning problems in IAFs to model checking problems in DL-PA. 

\begin{proposition}\label{prop:rediafs} 
Let $\niaf=\iaf$, $\sigma \in \semlist$, and $a \in \argset^{\fix}$. Then:
\begin{itemize}
\item The answer to $\sigma$-NSA with input $\niaf$ and $a$ is yes iff\\
 $v_{\niaf}\models [\makecomp^{\niaf};\mkext^{\sigma}] \acc a$.
\item The answer to $\sigma$-NCA with input $\niaf$ and $a$ is yes iff \\ $v_{\niaf}\models [\makecomp^{\niaf}]\langle\mkext^{\sigma}\rangle \acc a$.
\item The answer to $\sigma$-PCA with input $\niaf$ and $a$ is yes iff \\ $v_{\niaf}\models \langle\makecomp^{\niaf};\mkext^{\sigma}\rangle \acc a$.
\item The answer to $\sigma$-PSA with input $\niaf$ and $a$ is yes iff \\ $v_{\niaf}\models \langle\makecomp^{\niaf}\rangle [\mkext^{\sigma}] \acc a$.
\end{itemize}
\end{proposition}

\subsection{Rich Incomplete AFs}\label{sec:riafs}
A \textbf{rich incomplete AF} (rIAF) \cite{mailly2020note} extends an IAF in its uncertain part $\unc$ by adding a new (symmetric and irreflexive) uncertainty relation 
$\symattrel \subseteq (\argset^{\fix}\cup \argset^{?})\times (\argset^{\fix}\cup \argset^{?})$ 
such that $\symattrel \cap \attrel^{\fix} = \emptyset$ and $\symattrel \cap \attrel^{?} = \emptyset$.
We sometimes omit internal brackets when talking about rIAFs and note them $\riaf$. The new component, $\symattrel$, is informally understood as a set of attacks whose existence is known, but whose direction is unknown. The introduction of $\symattrel$ can be motivated by pointing out that attacks have two essential properties: their existence and their direction. Thus, while $\attrel^{?}$ captures uncertainty about the former, $\symattrel$ captures uncertainty about the latter. Note that any IAF can be understood as a rIAF with empty $\symattrel$. 
The notion of completion is easily adapted to rIAFs, capturing the intuitions about $\symattrel$ that we have just mentioned. A \textbf{completion} of $\nriaf=\riaf$ is any AF $\comp$ such that:
\begin{itemize}
\item 
$\argsf \subseteq \argset^{\ast}\subseteq (\argsf\cup\argset^{?})$; 
\item 
$\attfix\upharpoonright_{ \argset^{\ast}}\subseteq \attrel^{\ast}\subseteq (\attfix\cup \attrel^{?}\cup \symattrel)\upharpoonright_{ \argset^{\ast}}$; 
\item 
for every $x,y\in \argset^{\ast}$: $(x,y)\in \symattrel$ implies $(x,y)\in \attrel^{\ast}$ or $(y,x)\in \attrel^{\ast}$.
\end{itemize}
  
\begin{example}\label{ex:riaf}
Let $\nriaf_0=(\argset_0^{\fix},\argset_0^{?},\attrel^{\fix}_0,\attrel_0^{?},\attrel_0^{\leftrightarrow})$ where $\argset^{\fix}_0=\{a,b,d\}$, $\argset_0^{?}=\{c,e,f\}$, $\attrel_{0}^{\fix}=\{(b,a), (d,a), (c,b), (e,d), (f,e) \}$, $\attrel_{0}^{?}=\{(f,c) \}$, and $\attrel_{0}^{\leftrightarrow}=\{(c,e),(e,c)\}$. 
We represent $\nriaf_0$ graphically as follows:
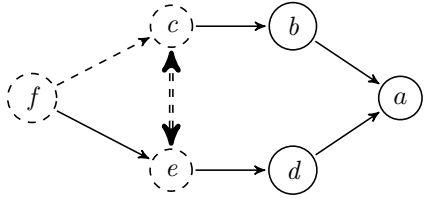
\begin{center}
\begin{tikzpicture}[modal,world/.append style=
{minimum size=0.5cm}]
\node[world] (a) [] {$a$};
\node (pos) [left=1 cm of a]{};
\node[world] (b) [above=0.5 cm of pos]{b};
\node[world,dashed] (c) [left=1 cm of b]{c};
\node[world] (d) [below=0.5 cm of pos]{d};
\node[world,dashed] (e) [left=1 cm of d]{e};
\node[world,dashed] (f) [left=3 cm of pos]{f};
\draw[->] (b) edge (a);
\draw[->] (d) edge (a);
\draw[->] (c) edge (b);
\draw[->] (e) edge (d);
\draw[<->] (c) edge[double,dashed] (e);
\draw[->] (f) edge (e);
\draw[->] (f) edge[dashed] (c);
\end{tikzpicture}
\end{center}
The set of completions of $\nriaf_0$ is depicted in Table~\ref{tab:comp}. 

\end{example}
 
\medskip
The computation of the completions of a rich IAF in DL-PA gets slightly more complicated since the program $\mktruesome$ does not suffice to deal with the symmetric attacks of $\symattrel$.
We can, however, define a specific program for this purpose.

First of all, given $\nriaf=\riaf$, the \textbf{valuation associated to $\nriaf$} is determined by its fixed part as before:

\begin{align*}
v_{\nriaf}
&=v_{(\argset^{\fix},\attrel^{\fix})} 
\\&= \awset_{\argset^{\fix}} \cup \attvarset_{\attrel^{\fix}}
\\&= \{\aw x \mid x \in \argsf\}\cup\{\att x y \mid (x,y) \in \attfix\} .
\end{align*}

Note that, contrarily to what happened with IAFs, $(\argset_{\val_{\nriaf}},\attrel_{\val_{\nriaf}})$ is \emph{not} always a completion of $\nriaf$: this fails to be the case as soon as $\symattrel \cap (\argset^{\fix}\times \argset^{\fix})$ is nonempty.
Let us now define the program that integrates the elements of $\symattrel$ into each completion.

Let 
$\attvarset_{\attrel}=\{\att{x_{1}}{y_{1}},...,\att{x_{n}}{y_{n}} \}$
\footnote{Remember that  
$\attvarset_{\attrel}= \{\att x y \mid (x,y) \in \attrel\}$, and that $\attvarset_{\attrel}$ is a subset of the set of propositional variables $\attvarset_{\uni \times \uni}$.} 
be a set of attack variables, and define the program
\begin{align*}
\dis(\attvarset_{\attrel})=\left( \assgntop{ \att{x_{1}}{y_{1}} } \cup \assgntop{\att{y_{1}}{x_{1}} } \right); \ldots ;\left( \assgntop{ \att{x_{n}}{y_{n}} } \cup \assgntop{ \att{y_{n}}{x_{n}} } \right)\text{.}
\end{align*}
Intuitively, $\dis(\attvarset_{\attrel})$ makes true at least one of the variables from the set $\{\att x y, \att y x\}$, for each $(x,y)\in \attrel$. 
Moreover, when applied to a symmetric relation $\symattrel$, $\dis$ makes true either $\att x y$, or $\att y x$, or both, for every $(x,y)\in \symattrel$.

We are now ready to define the program $\makecomp$ in its version for rIAFs. 
Given $\nriaf=\riaf$, let 
\begin{align*}
\makecomp^{\nriaf}&= \mktruesome(\awset_{\argset^{?}});\mktruesome(\attvarset_{\attrel^{?}});\dis(\attvarset_{\symattrel})\text{.}
\end{align*}
The following proposition states that the above program is correct.
\begin{proposition}\label{prop:riafenco} Let $\nriaf=\riaf$, then: 
\begin{itemize}
\item If $(\val_{\nriaf},\val)\in ||\makecomp^{\nriaf}||$, then $(\argset_{\val},\attrel_{\val})\in \mathsf{completions}(\nriaf)$.
\item If $(\argset^{\ast},\attrel^{\ast})\in \mathsf{completions}(\nriaf)$, then $(\val_{\nriaf},\val_{(\argset^{\ast},\attrel^{\ast})})\in ||\makecomp^{\nriaf}||$.
\end{itemize} \end{proposition}

Again, acceptance problems can be reduced to DL-PA model checking problems. Note that the definition of acceptance problems for rIAFs is just as for IAFs (we only have to change the input). Let us just state the reduction result we are after:

\begin{proposition}\label{prop:redriafs} 
Let $\sigma \in \semlist$. Let $\nriaf=\riaf$ and $a \in \argset^{\fix}$. Then:

\begin{itemize}

\item The answer to $\sigma$-NSA with input $\nriaf$ and $a$ is yes iff\\
 $v_{\nriaf}\models [\makecomp^{\nriaf};\mkext^{\sigma}] \acc a$.
\item The answer to $\sigma$-NCA with input $\nriaf$ and $a$ is yes iff \\ $v_{\nriaf}\models [\makecomp^{\nriaf}]\langle\mkext^{\sigma}\rangle \acc a$.
\item The answer to $\sigma$-PCA with input $\nriaf$ and $a$ is yes iff \\ $v_{\nriaf}\models \langle\makecomp^{\nriaf};\mkext^{\sigma}\rangle \acc a$.
\item The answer to $\sigma$-PSA with input $\nriaf$ and $a$ is yes iff \\ $v_{\nriaf}\models \langle\makecomp^{\nriaf}\rangle [\mkext^{\sigma}] \acc a$.

\end{itemize}
\end{proposition}

\subsection{Shrinking the Set of Completions}

Incomplete AFs (and their enriched version) deal with uncertainty about argumentative situations in a simple and intuitive manner. However, the kind of situations that we can model with them is rather limited (as we will discuss in detail later on). This is the main motivation for the development of more expressive formalisms, and it actually led to concurrent proposals during the last year, either under the name of \emph{constrained incomplete argumentation frameworks} \cite{clar2021,maillyciafs} or \emph{incomplete argumentation frameworks with dependencies} \cite{fazzingaijcai21,fazzingakr21}. We start by presenting our version of constrained incomplete AFs (the one introduced in \cite{clar2021}), and then move to alternative approaches.

\subsubsection{Constrained Incomplete AFs}\label{sec:ciafs}
A \textbf{constrained incomplete AF} (cIAF) is a pair $\nciaf=(\argset,\varphi)$ where $\argset\subseteq \uni$ is a set of arguments and $\varphi$ is a Boolean formula built over the set of propositional variables $\awset_{\argset}\cup \attvarset_{\argset\times \argset}$.\footnote{We have slightly changed the original definition of cIAFs \cite{clar2021}, by switching the domain from $\uni$ to an arbitrary $\argset$, because it allows for naturally plugging-in argumentation dynamics, as we will do in Section~\ref{sec:cciafs}.}
The set of \textbf{completions} of a given cIAF is 
$$\mathsf{completions}(\argset,\varphi) = \{(\argset_\val,\attrel_\val)\mid \val \subseteq \propset_{\argset} \text{ and } \val \models \varphi\} . $$

\begin{example}\label{ex:ciaf} Let us consider $\nciaf_0 = ({\argset},\varphi)$ with ${\argset}=\{a,b\}$ and $\varphi=(\aw a \land \aw b) \land (\att a b \lor \att b a) \land \lnot (\att a b \land \att b a) \land \lnot \att a a \land \lnot \att b b$. The completions of $\nciaf_0$ are:
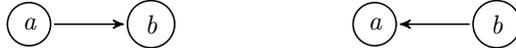
\begin{center}
\begin{tikzpicture}[modal,world/.append style=
{minimum size=0.5cm}]
\node[world] (a) []{a};
\node[world] (b) [right=1 cm of a]{b};
\draw[->] (a) edge (b);

\node[world] (a1) [right=4 cm of a]{a};
\node[world] (b1) [right=1 cm of a1]{b};
\draw[->] (b1) edge (a1);

\end{tikzpicture}
\end{center}
\end{example}

Notice that, differently to what happened with previous classes of structures, the set of completions of a cIAF might be empty, since $\varphi$ can be an inconsistent formula. Moreover, even being consistent, $\varphi$ might not be satisfied by any valuation representing a non-empty AF, so that we could get the empty AF $(\emptyset,\emptyset)$ as the only completion of a cIAF; e.g., $\compfunc((\{a\},\lnot \aw a))=\{(\emptyset,\emptyset)\}$.

\paragraph{The need of cIAFs.} 
Besides being mathematically interesting, one may wonder why one should use cIAFs. 
As mentioned, our main motivation is that, while the computational complexity of reasoning tasks associated to the previously introduced formalisms (i.e., (r)IAFs and subclasses) is well-known and relatively low, their modelling power is rather limited. Consider, for instance, a proponent reasoning about the view of her opponent in a very simple debate containing only two arguments $\{a,b\}$. Suppose that $a$ is an argument about public health policies stated by the right-wing presidential candidate. Similarly, $b$ is an argument stated by the left-wing candidate. Imagine that $a$ and $b$ have contradictory conclusions, so they are mutually incompatible. Let us informally understand $\attrel$ as a \emph{defeat} relation here, that is, a relation based on logical incompatibility plus some kind of epistemic-based assessment of the involved arguments (for instance, regarding the reliability of their premisses), as it is usually done in structured argumentation. Now, suppose our proponent knows that her opponent is polarized, in the sense that he (the opponent) is already inclined towards one side of the political spectrum, but she does not know which one; then the possible AFs that the agent attributes to her opponent are exactly the completions of $\nciaf_0$ (see Example~\ref{ex:ciaf}). As it will be proved later (Proposition~\ref{prop:express}), there is no rIAF (and therefore no IAF) with the exact set of completions of $\nciaf_0$.

\medskip

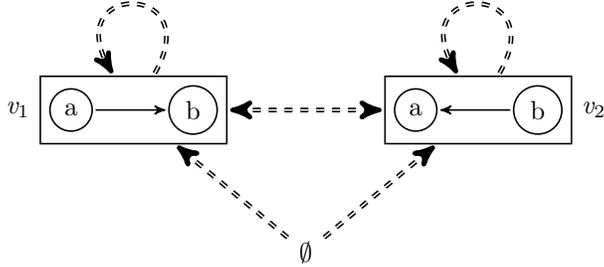
\begin{figure}
\centering
\begin{tikzpicture}[modal,world/.append style=
{minimum size=0.5cm}]

\node[world] (a) []{a};
\node[world] (b) [right=1 cm of a]{b};
\draw[->] (a) edge (b);
\node[draw, fit=(a) (b)](fit) [label=left:$\val_1$] {};

\node[world] (a1) [right=4 cm of a]{a};
\node[world] (b1) [right=1 cm of a1]{b};
\draw[->] (b1) edge (a1);
\node[draw, fit=(a1) (b1)](fit1) [label=right:$\val_2$] {};

\node (pos) at ($(fit)!0.5!(fit1)$) {};
\node (emp) [below=1.5cm of pos] {$\emptyset$};

\draw[->] (emp) edge[double,dashed] (fit);
\draw[->] (emp) edge[double,dashed] (fit1);
\draw[->] (fit1) edge[reflexive above, double,dashed] (fit1);
\draw[->] (fit) edge[reflexive above, double,dashed] (fit);
\draw[<->] (fit) edge[double,dashed] (fit1);
\end{tikzpicture}
\caption{Completions of $\nciaf_0$ seen as valuations over $\propset_{\{a,b\}}$. Dashed double arrows represent the interpretation of $\makecomp^{\nciaf_{0}}$; the other valuations over $\propset_{\{a,b\}}$ are omitted.}
\label{fig:ciafasmodel}
\end{figure}

Let us now show how cIAFs can be captured in DL-PA. Let $\nciaf=\ciaf$, and define its \textbf{associated valuation} simply as the empty set, that is, $\val_{\nciaf}=\emptyset$.
(Actually any valuation over $\propset_{\argset}$ will do the job.) 
The program generating all completions of $\nciaf$ is defined as
$$\makecomp^{\nciaf}=\vary(\awset_{\argset});\vary(\attvarset_{\argset \times \argset});\varphi?\text{.}$$
The behaviour of $\makecomp^{\nciaf_0}$ (see Example \ref{ex:ciaf}) is illustrated in Figure~\ref{fig:ciafasmodel}. 

\black
\begin{proposition}\label{prop:encociafs} Let $\nciaf=\ciaf$, then: 
\begin{itemize}
\item If $(\val_{\nciaf},\val)\in ||\makecomp^{\nciaf}||$, then $(\argset_{\val},\attrel_{\val})\in \mathsf{completions}(\nciaf)$.
\item If $(\argset^{\ast},\attrel^{\ast})\in \mathsf{completions}(\nciaf)$, then $(\val_{\nciaf},\val_{(\argset^{\ast},\attrel^{\ast})})\in ||\makecomp^{\nciaf}||$.
\end{itemize} \end{proposition}

\black

Reasoning problems for (r)IAFs can be easily adapted to cIAFs: we just have to ensure that the argument about which we formulate the query belongs to all completions.
As an example, consider:
\begin{center}
\begin{tabular}{|l|}
\hline
 $\sigma$-Necessary-Credulous-Acceptance ($\sigma$-NCA)\\
\hline
\textbf{Given:}  A constrained IAF
 $\nciaf=\ciaf$ \\ and an argument $a\in \argset$ such that $\models \varphi \to \aw a$. \\

\textbf{Question:}  Is it true that for every \\ $(\argset^{\ast},\attrel^{\ast}) \in \mathsf{completions}(\nciaf)$ \\ there is an $E\in \sigma(\argset^{\ast},\attrel^{\ast})$ such that $a \in E$? \\
\hline
\end{tabular}
\end{center}

\noindent Note that requiring $\models \varphi \to \aw a$ amounts to requiring $a \in \argset$ for all $(\argset,\attrel)\in \mathsf{completions}(\argset,\varphi)$. 
 
Once again, we can reduce acceptance problems in cIAFs to DL-PA model checking problems.

\begin{proposition}\label{prop:redciaf} 
Let $\nciaf=\ciaf$ and let $a \in \argset$ such that $\models \varphi \to \aw a$. Let $\sigma \in \semlist$. Then:

\begin{itemize}

\item The answer to $\sigma$-NSA with input $\nciaf$ and $a$ is yes iff\\
 $v_{\nciaf}\models [\makecomp^{\nciaf};\mkext^{\sigma}] \acc a$.
\item The answer to $\sigma$-NCA with input $\nciaf$ and $a$ is yes iff \\ $v_{\nciaf}\models [\makecomp^{\nciaf}]\langle\mkext^{\sigma}\rangle \acc a$.
\item The answer to $\sigma$-PCA with input $\nciaf$ and $a$ is yes iff \\ $v_{\nciaf}\models \langle\makecomp^{\nciaf};\mkext^{\sigma}\rangle \acc a$.
\item The answer to $\sigma$-PSA with input $\nciaf$ and $a$ is yes iff \\ $v_{\nciaf}\models \langle\makecomp^{\nciaf}\rangle [\mkext^{\sigma}] \acc a$.
\end{itemize}
\end{proposition}

We observe that beyond these reasoning problems one may also consider the reasoning task of checking emptiness of the set of completions of a cIAF.

\subsubsection{Closely Related Approaches}\label{sec:related}
As mentioned, the idea of shrinking the set of completions of an IAF led to concurrent proposals during the last year. In this subsection, we briefly present the two alternative approaches to our cIAFs of \cite{clar2021}.  
 
\paragraph{A more graph-theoretic version of cIAFs.}
In \cite{maillyciafs}, Jean-Guy Mailly defined his version of cIAFs that we call \textbf{cIAFs$^{JM}$} here to avoid confusion. A cIAF$^{JM}$ is pair of the form $(\niaf,\varphi)$ where $\niaf=\unraveliaf$ is an IAF and $\varphi$ is a Boolean formula over $\propset^{\niaf}=\awset_{\argset\cup \argset^{?}}\cup \attvarset_{(\argset\cup \argset^{?}) \times (\argset\cup \argset^{?})}$. Then the set of \textbf{completions of} $\nciaf^{JM}=(\niaf,\varphi)$ is defined as 
$$\compfunc(\niaf)\cap\{(\argset_\val,\attrel_\val)\mid \val \subseteq \propset^{\niaf} \text{ and } v\models \varphi\}\text{.}$$

\paragraph{IAFs with dependencies.} In \cite{fazzingaijcai21,fazzingakr21}, the team from the University of Calabria formed by Bettina Fazzinga, Sergio Flesca and Filippo Furfaro introduced the notion of IAFs with dependencies.\footnote{\label{correlation}The term \emph{correlations} is used in \cite{fazzingaijcai21,fazzingakr21} as the informal counterpart of \emph{dependencies}. We stick to the latter term to avoid confusion.} More precisely, their two proposals respectively focus on two restricted classes of IAFs that we have already mentioned: arg-IAFs, and att-IAFs. For the sake of brevity we only present here the notion of arg-IAF with dependencies of \cite{fazzingaijcai21}. 
Let $\argset$ be a set of arguments and let $X,Y\subseteq \argset$. 
First, a \textbf{dependency over $\argset$} is either $X \Rightarrow Y$ or $\mathsf{OP}(X)$ with $\mathsf{OP}\in\{\mathsf{OR}, \mathsf{NAND}, \mathsf{CHOICE}\}$. 
Second, an \textbf{arg-IAF with dependencies} (d-arg-IAF, for short)
is a pair $((\argset,\argset^{?},\attrel),\Delta)$, where $(\argset,\argset^{?},\attrel)$ is an arg-IAF and $\Delta$ is a set of dependencies over $\argset^{?}$. Before defining the completions of a d-arg-IAF we need to settle how dependencies 
are to be interpreted in arg-IAFs. 
Let $\af$ be an AF and let $\delta$ be a dependency over $\argset$. 
We say that $\af$ satisfies $\delta$\footnote{
\cite{fazzingaijcai21} uses the expression ``$\af$ is valid w.r.t.\ $\delta$'', but our expression is more appropriate in a logical analysis.}
iff one of the following mutually exclusive clauses holds:

\begin{itemize}
\item $\delta=X \Rightarrow Y$ and (if $X\subseteq \argset$, then $\argset\cap Y\neq \emptyset$);
\item $\delta=\mathsf{OR}(X)$ and $\argset\cap X\neq \emptyset$, 
\item $\delta=\mathsf{NAND}(X)$ and $\argset\cap X\subset X$, 
\item $\delta=\mathsf{CHOICE}(X)$ and $|\argset \cap X|=1$. 
\end{itemize}

The \textbf{completions of $((\argset,\argset^{?},\attrel),\Delta)$} are defined as those completions of the arg-IAF 
$(\argset,\argset^{?},\attrel)$ that satisfy every dependency $\delta \in \Delta$.

\medskip
The three alternative proposals are already compared in \cite{mailly2021yes}. We will provide some new insights beyond this in the next section. Let us just make a couple of points here. 
First, note that both versions of cIAFs as well as IAFs with dependencies are clearly inspired by the notion of \emph{constrained AF} \cite{constrained}, which are pairs $(\af,\varphi)$ where $\varphi$ is used to shrink the set of \emph{extensions} of $\af$. Second, note that the reasoning tasks associated to both classes of structures are clearly encodable in DL-PA, but we do not work out the details here. Let us just point out that each set of dependencies $\Delta$ can be translated into a Boolean formula $t(\Delta)$, and then the program
$\makecomp^{((\argset,\argset^{?},\attrel),\Delta)}=\vary(\awset_{\argset^{?}});t(\Delta)?$  computes all the completions of $((\argset,\argset^{?},\attrel),\Delta)$ when executed at $\val_{(\argset,\attrel)}$.

\subsection{Comparison of the Different Approaches}\label{sec:comparison}

Let us now compare the different approaches to representing qualitative uncertainty about AFs. We start with a couple of general considerations.

\paragraph{Combinatorics vs.\ logic.} The spirit of the seminal works on IAFs was to represent uncertainty by defining completions as directed graphs whose domains and relations fall between given intervals. One may qualify this approach as ``combinatorial'', since, once the extremes of the interval are given (e.g.\ $\argset^{\fix}$ and $\argset^{\fix}\cup \argset^{?}$), the task of computing completions amounts to finding all possible combinations 
within the interval. Progressively, other reasoning features that we might qualify as ``logical'' have been integrated in the definition of completion. For instance, rIAFs introduce a sort of disjunctive reasoning through the addition of $\symattrel$. We can understand this transition from combinatorics to logic as a sort of spectrum: 

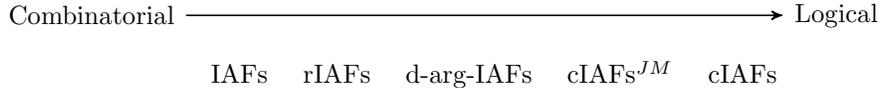
\begin{center}
\begin{tikzpicture}[modal,world/.append style=
{minimum size=0.5cm}]
\node (com) {Combinatorial};
\node (log) [right=8 cm of com] {Logical};

\node (pos) [right=4 cm of com] {};
\node (spe) [below=0.35 cm of pos]{IAFs \quad rIAFs \quad d-arg-IAFs \quad cIAFs$^{JM}$ \quad cIAFs };
\draw[->] (com) edge (log);
\end{tikzpicture}
\end{center}
Note how at the right-hand extreme (cIAFs), the combinatorial nature of completions has completely vanished.

\paragraph{Graphic representations.} One of the appealing features of IAFs is that they admit a very intuitive graphic representation (see Example~\ref{ex:iaf}). Interestingly, rIAFs and d-arg-IAFs can also be fully represented in a pictorial manner; see \cite{ijcai21} for examples with d-arg-IAFs. As pointed out in \cite{mailly2020note}, cIAFs$^{JM}$ only admit a partial graphic representation. Finally, this pictorial representability is lost by our cIAFs, which completely abstract away from the graph-theoretic definition of IAFs. Hence, in this respect, IAFs with dependencies compare better to cIAFs and cIAFs$^{JM}$. 
 \paragraph{Expressivity via sets of completions.} 
 Following \cite{mailly2020note}, we can compare the modelling power of each of the previous formalisms for arguing with uncertainty using the sets of completions they can represent.
Let $\mathcal{IAF}$ denote the class of all IAFs, and likewise for 
$att\text{-}\mathcal{IAF}$, $arg\text{-}\mathcal{IAF}$, $\mathcal{RIAF}$, $d\text{-}arg\text{-}\mathcal{IAF}$, $d\text{-}att\text{-}\mathcal{IAF}$, $c\text{-}\mathcal{IAF}$ and $c\text{-}\mathcal{IAF}^{JM}$.
Let $\mathcal{X}$ and $\mathcal{Y}$ be metavariables denoting one of these classes.
We say that $\mathcal{X}$ is \textbf{at least as expressive as}
$\mathcal{Y}$ (in symbols: $\mathcal{X}\succeq\mathcal{Y}$) if, for every $Y \in \mathcal{Y}$ there is a $X \in \mathcal{X}$ such that $\mathsf{completions}(X)=\mathsf{completions}(Y)$. 
We use $\succ$ to denote the strict part of $\succeq$,
we use $\preceq$ to denote the inverse of $\succeq$, and we use $\equiv$ to abbreviate $\succeq\cap \preceq$. For instance, it was proved in \cite{mailly2020note} that $\mathcal{RIAF} \succ \mathcal{IAF}$. 

\begin{proposition}\label{prop:express}
cIAFs are strictly more expressive than IAFs and rIAFs. In other words, for every (r)IAF, there is a cIAF with the same set of completions; but there is a cIAF such that no (r)IAF has the same set of completions.  \end{proposition}

In the first part of the proof of the previous proposition---see the \nameref{sec:app}---we have used an argument that works for \emph{any} set of directed graphs with domain $\uni$ (and not only for the completions of a given rIAF), hence we can state that:

\begin{corollary} For any set $\mathsf{S}$ of directed graphs with domain $\uni$ there is a cIAF $\nciaf$ such that $\mathsf{S}=\mathsf{completions}(\nciaf)$. \end{corollary}
\black

\begin{figure}
\centering

\begin{tikzpicture}[modal,world/.append style=
{minimum size=0.5cm}]
\node (afs) []{$\mathcal{AF}$};
\node (pos) [right=0.5 cm of afs]{};
\node (attiafs) [above=0.75 cm of pos]{$att\text{-}\mathcal{IAF}$};

\node (dattiafs) [right=0.75 cm of attiafs]{$d\text{-}att\text{-}\mathcal{IAF}$};

\node (argiafs) [below=0.75 cm of pos]{$arg \text{-}\mathcal{IAF}$};

\node (dargiafs) [right=0.75 cm of argiafs]{$d\text{-}arg\text{-}\mathcal{IAF}$};
\node (iafs) [right=0.75 cm of pos]{$\mathcal{IAF}$};
\node (pos1) [right=0.45 cm of iafs]{};

\node (riafs) [right=0.75 cm of iafs] {$\mathcal{RIAF}$};

\node (ciafs) [right=0.75 cm of riafs]{$c\text{-}\mathcal{IAF}\text{,}c\text{-}\mathcal{IAF}^{JM}$};

\draw[->] (afs) edge (attiafs);
\draw[->] (afs) edge (argiafs);
\draw[->] (attiafs) edge (iafs);
\draw[->] (attiafs) edge (dattiafs);
\draw[->] (argiafs) edge (iafs);
\draw[->] (argiafs) edge (dargiafs);
\draw[->] (iafs) edge (riafs);
\draw[->] (dargiafs) edge (ciafs);
\draw[->] (dattiafs) edge (ciafs);
\draw[->] (riafs) edge (ciafs);

\end{tikzpicture}

\caption{Relative expressivity of formalisms for qualitative uncertainty in formal argumentation. An arrow from $\mathcal{X}$ to $\mathcal{Y}$ means that $ \mathcal{X} \preceq \mathcal{Y}$, i.e., $\mathcal{Y}$ is at least as expressive as $\mathcal{X}$. Reflexive and transitive arrows have been omitted.}
\label{fig:exp}
\end{figure}
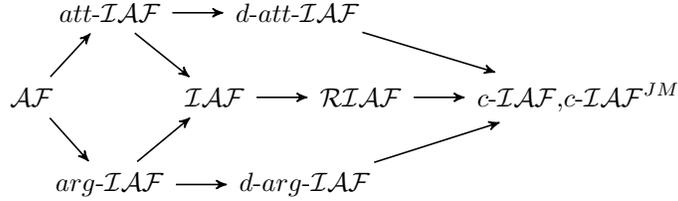

In words, cIAFs are a maximally expressive formalism for representing qualitative uncertainty about AFs. Using arguments similar to those employed in the proof of Proposition~\ref{prop:express} we can provide the following general result:

\begin{proposition}\label{prop:ex}
The relations of Figure~\ref{fig:exp} hold, where an arrow from $\mathcal{X}$ to $\mathcal{Y}$ means that $ \mathcal{X} \preceq \mathcal{Y}$ and where transitive and reflexive arrows are omitted.  
\end{proposition}

Besides providing a full expressivity map, this proposition highlights the fact that IAFs with dependencies have not been given their most expressive formulation yet. That is, we have arg-IAFs with dependencies \cite{fazzingaijcai21}, and att-IAFs with dependencies \cite{fazzingakr21}, but no IAFs with dependencies. This makes that these kinds of structures do not yet permit expressing any set of completions (contrarily to what happens with both cIAFs and cIAFs$^{JM}$). 
It seems clear that a mixed version of those formalisms would also be maximally expressive. However, some important design choices are to be made; for instance, whether one permits \emph{mixed dependencies} (those involving uncertain arguments and attacks) or not.

\section{Encompassing Dynamics and Uncertainty}\label{sec:dynamics}

 As argued in the introduction, there are two fundamental aspects of argumentation that are left out of AFs: the uncertainty about the relevant argumentative information (that is, which arguments and attacks should be taken into account during a debate), and the dynamics of such information. In the previous section we have discussed various ways to represent uncertainty about AFs. As to the dynamics of AFs, it is a well-studied branch of research by now; see e.g.\ \cite{DM18,baumann2021enforcement} for recent surveys. In this section we sketch
how both ideas are to be combined. We start by presenting a well-studied case: control AFs \cite{dimopoulos2018control}, showing that their main reasoning tasks are also encodable in DL-PA. After mentioning some of its limitations, we proceed to study an extension that combines the kind of dynamics captured by CAFs with the flexibility of cIAFs for representing uncertainty. We close the section by sketching a general theory of dynamics and uncertainty of AFs that provides conceptual tools for conducting future research.

\subsection{Control AFs}\label{sec:cafs}
Control argumentation frameworks were introduced in \cite{dimopoulos2018control} and applied to argument-based negotiation in \cite{cafsnegotiation}. They represent a joint approach to uncertainty and dynamics of AFs. Regarding uncertainty, they are as expressive as rIAFs (Section~\ref{sec:riafs}). As to dynamics, they capture a parametrised version of what has been called \emph{normal expansion} \cite{BB10} at the level of each completion. 
 
Formally, a \textbf{control argumentation framework} is a triple $\ncaf= \caf$ where:
\begin{itemize}
\item 
$\fix=(\argsf,\attfix)$ is the \emph{fixed part}, with $\attfix\subseteq (\argsf\cup  \argset^{?})\times (\argsf\cup\argset^{?})$, and both $\argsf$ and $\argset^{?}$ being two finite sets of arguments;
\item 
$\unc=(\argset^{?},(\attrel^{?}\cup \symattrel))$ is the \emph{uncertain part}, where $$\attrel^{?},\symattrel \subseteq (\argsf\cup \argset^{?})\times(\argsf\cup\argset^{?})$$ and $\symattrel$ is symmetric and irreflexive;\footnote{Symmetry and irreflexivity of $\symattrel$ are not assumed in the original paper \cite{dimopoulos2018control}, but appeared later on in the literature about CAFs \cite{niskanen2021controllability,niskanen2020thesis}.
Note that both assumptions do not affect expressivity (in the sense used in Section~\ref{sec:comparison}) of CAFs.}
\item 
$\con=(\argset^{\con},\attrel^{\con})$ is the \emph{control part}, where $\argset^{\con}$ is yet another finite set of arguments and $$\attrel^{\con}\subseteq (\argset^{\con}\times (\argset^{\fix} \cup \argset^{?}\cup \argset^{\con})) \cup ( (\argset^{\fix} \cup \argset^{?}\cup \argset^{\con})\times \argset^{\con} )\text{;}$$
\item $\argsf$, $\argset^{?}$, and $\argset^{\con}$ are pairwise disjoint; and
\item $\attfix,\attrel^{?},\symattrel$, and $ \attrel^{\con}$ are pairwise disjoint.
\end{itemize}
We note $\mathcal{CAF}$ the class of all control AFs.
 
Given a $\ncaf=\caf$, a \textbf{control configuration} is a subset of control arguments $\cfg \subseteq \argset^{\con}$. Informally, each control configuration can be seen as a possible argumentative move for the proponent.
The \textbf{CAF associated to $\cfg$} 
is $\ncaf_{\cfg} = (\fix,\con_{\cfg},\unc)$, where 
 $\con_{\cfg} = (\cfg,\attrel^{\con}\upharpoonright_{ \argsf\cup \argset^{?}\cup \cfg})$.

\begin{example}\label{ex:caf} 
Let us consider the CAF $\ncaf_0 = (\fix_0,\con_0,\unc_0)$ 
where $\argset^{\fix}_0=\{a\}$, $\attrel_0^{\fix}=\{(f,e)\}$, $\argset^{?}_{0}=\{c,e,f\}$, $\attrel^{?}_0=\{(f,c)\}$, $\attrel^{\leftrightarrow}_0=\{(c,e),(e,c)\}$, $\argset^{\con}_{0}=\{b,d\}$, and $\attrel^{\con}_{0}=\{(b,a),(d,a),(c,b),(e,d)\}$.
We represent $\ncaf_0$ graphically as follows:
 
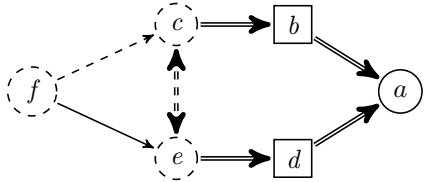
\begin{center}
\begin{tikzpicture}[modal,world/.append style=
{minimum size=0.5cm}]
\node[world] (a) [] {$a$};
\node (pos) [left=1 cm of a]{};
\node[carg] (b) [above=0.5 cm of pos]{b};
\node[world,dashed] (c) [left=1 cm of b]{c};
\node[carg] (d) [below=0.5 cm of pos]{d};
\node[world,dashed] (e) [left=1 cm of d]{e};
\node[world,dashed] (f) [left=3 cm of pos]{f};
\draw[->] (b) edge[double] (a);
\draw[->] (d) edge[double] (a);
\draw[->] (c) edge[double] (b);
\draw[->] (e) edge[double] (d);
\draw[<->] (c) edge[double,dashed] (e);
\draw[->] (f) edge (e);
\draw[->] (f) edge[dashed] (c);
\end{tikzpicture}
\end{center}
\end{example}

The notion of \textbf{completion} is defined as follows for CAFs:
\begin{itemize}
\item 
$(\argsf\cup \argset^{\con})\subseteq \argset^{\ast}\subseteq (\argsf\cup \argset^{\con}\cup \argset^{?})$;
\item 
$(\attfix\cup\attrel^{\con})\upharpoonright_{ \argset^{\ast}}\subseteq \attrel^{\ast}\subseteq (\attfix\cup \attrel^{\con}\cup\attrel^{?}\cup \symattrel)\upharpoonright_{ \argset^{\ast}}$; and
\item 
for every $x,y\in \argset^{\ast}$: $(x,y)\in \symattrel$ implies $(x,y)\in \attrel^{\ast}$ or $(y,x)\in \attrel^{\ast}$.
\end{itemize}    

According to this definition, control arguments/attacks behave like fixed arguments/attacks once they have been communicated. Hence, the completions of $\ncaf_0$ coincide with those of $\nriaf_0$ (Example \ref{ex:riaf}), i.e., those depicted in Table~\ref{tab:comp}. 

Regarding CAFs, defining relevant reasoning tasks gets slightly more complicated because we have to take into account their dynamic dimension. In this context, a natural reasoning task is to find a control configuration (that is, a set of control arguments) such that a certain argument gets accepted by the opponent after the latter learns about them. 
As before, acceptability is then relative to quantification over completions and extensions. 
Here is an example: 

\begin{center}
\begin{tabular}{|l|}
\hline
 $\sigma$-Necessary-Sceptical-Controllability ($\sigma$-NSCon)\\
\hline

\textbf{Given:} A control argumentation framework  \\
 $\ncaf=\caf$ and an argument $a\in \argsf$. \\

\textbf{Question:} Is it true that there is a configuration \\ $\cfg\subseteq \argset_{\con}$ such that for every completion
$(\argset^{\ast},\attrel^{\ast})$  \\ of $\ncaf_{\cfg}$ and for every $E\in \sigma(\argset^{\ast},\attrel^{\ast}),a \in E$? \\
\hline
\end{tabular}
\end{center}

\medskip
We now move on to explain how to reason about CAFs in DL-PA. Since, uncertainty-wise, control argumentation frameworks are essentially rich incomplete argumentation frameworks, the delicate part in the encoding process comes with their dynamic component, i.e., the control part. 

First, given a CAF $\ncaf=\caf$, we define its \textbf{associated valuation} as
\begin{align*}
v_{\ncaf}
&=v_{(\argset^{\fix},\attrel^{\fix} \cup \attrel^{C})} 
\\&= \awset_{\argset^{\fix}} \cup \attvarset_{\attrel^{\fix}} \cup \attvarset_{\attrel^{C}}
\\&= \{\aw x \mid x \in \argsf\}\cup\{\att x y \mid (x,y) \in \attfix\} 
\cup\{\att x y \mid (x,y)\in \attrel^{C}\}\text{.}
\end{align*}
Note that $v_{\ncaf}$ contains all attack variables corresponding to control attacks, but none of them appear in $(\argset_{v_{\ncaf}},\attrel_{v_{\ncaf}})$ since none of the control arguments has been communicated yet. This highlights the fact that in an epistemic interpretation of CAFs, the proponent knows how the opponent will perceive the attack relations regarding all communicable arguments. 

To capture the dynamic component of $\ncaf$ we define the following program:
$$\control^{\ncaf}= \mktruesome(\awset_{\argset^{\con}})\text{.}$$
Intuitively, $\control^{\ncaf}$ nondeterministically chooses some of the possible control configurations of $\ncaf$, i.e., some subset of control arguments. 
 
Once we have computed some control configuration, we use the same program as for rIAFs in order to compute completions:
\begin{align*}
\makecomp^{\ncaf}&=\mktruesome(\awset_{\argset^{?}});\mktruesome(\attvarset_{\attrel^{?}});\dis(\attvarset_{\symattrel})\text{.}
\end{align*}
We again state a correctness result:

\begin{proposition}\label{prop:cafsenco}
Let $\ncaf=\caf$. 
\begin{itemize}
\item If $(\val_{\ncaf},\val)\in ||\control^{\ncaf};\makecomp^{\ncaf}||$ then there is a control configuration $\cfg\subseteq \argset^{C}$ and a completion $\comp$ of $\ncaf_{\cfg}$ such that $(\argset_{\val},\attrel_{\val})=\comp$.
\item For every control configuration $\cfg\subseteq \argset^{\con}$ and every $\comp \in \mathsf{completions}(\ncaf_{\cfg})$ there is a valuation $\val\in 2^{\propset}$  such that $(\val_{\ncaf},\val)\in ||\control^{\ncaf};\makecomp^{\ncaf}||$ and $(\argset_{\val},\attrel_{\val})=\comp$.
\end{itemize} \end{proposition}

We can then combine the previous programs with $\mathsf{makeExt}$ in order to reduce controllability problems to DL-PA model checking problems. 

\begin{proposition}\label{prop:redcaf} 
Let $\sigma \in \semlist$. Let $\ncaf=\caf$ and $a \in \argset^{\fix}$. Then:

\begin{itemize}

\item The answer to $\sigma$-NSCon with input $\ncaf$ and $a$ is yes iff\\
 $v_{\ncaf}\models \langle \control^{\ncaf}\rangle[\makecomp^{\ncaf};\mkext^{\sigma}] \acc a$.
\item The answer to $\sigma$-NCCon with input $\ncaf$ and $a$ is yes iff \\ $v_{\ncaf}\models \langle \control^{\ncaf}\rangle [\makecomp^{\ncaf}]\langle\mkext^{\sigma}\rangle \acc a$.
\item The answer to $\sigma$-PCCon with input $\ncaf$ and $a$ is yes iff \\ $v_{\ncaf}\models \langle\control^{\ncaf};\makecomp^{\ncaf};\mkext^{\sigma}\rangle \acc a$.
\item The answer to $\sigma$-PSCon with input $\ncaf$ and $a$ is yes iff \\ $v_{\ncaf}\models \langle\control^{\ncaf};\makecomp^{\ncaf}\rangle [\mkext^{\sigma}] \acc a$.

\end{itemize}

\end{proposition}

We close this section by highlighting and making precise two of the main modelling limitations of CAFs that we have already mentioned. 
Regarding uncertainty, they cannot go further than rIAFs. As to dynamics, the form of communication that they model assumes that uncertainty does not increase. More formally:
\begin{remark} $\mathcal{CAF} \equiv \mathcal{RIAF}$. 
\end{remark}
\begin{remark}\label{remark:compcaf} Let $\ncaf=\caf$ and let $\cfg,\cfg^{'}\subseteq \argset^{\con}$. Then 
$ | \mathsf{completions}(\ncaf_{\cfg}) | = | \mathsf{completions}(\ncaf_{\cfg^{'}}) | $. 
\end{remark}

\subsection{Control Constrained Incomplete AFs}\label{sec:cciafs}
One of the advantages of the approaches presented so far is that they can be freely combined. Moreover, the encoding of these formalisms in DL-PA can easily be extrapolated to combined classes of structures. In this subsection, we give evidence of such flexibility by mixing the kind of uncertainty modelled by cIAFs with the kind of dynamics modelled by CAFs. Formally, a \textbf{control constrained incomplete AF} (CcIAF) is a tuple $\ncciaf=(\argset^{C}, \attrel^{C}, \argset^{S}, \varphi)$ with:
\begin{itemize}
\item $\argset^{C}$ (control arguments) and $\argset^{S}$ (static arguments) are disjoint,
\item $\attrel^{C} \subseteq (\argset^{C} \times(\argset^{C} \cup \argset^{S}))\cup ((\argset^{C}\cup \argset^{S})\times \argset^{C})$,
\item $\varphi$ is a Boolean formula over $\awset_{\argset^{S}}\cup\attvarset_{\argset^{S} \times \argset^{S}}$.
\end{itemize}
Given $\ncciaf=\cciaf$, the pair $(\argset^{S},\varphi)$ is its \textbf{underlying cIAF}. 
 
The notion of completion is adapted to CcIAFs by combining the intuition behind CAFs and cIAFs, i.e., a completion of $\cciaf$ is any AF $(\argset^{\ast},\attrel^{\ast})$ such that:

\begin{itemize}
\item $\argset^{\ast}= \argset^{C} \cup \argset'$,
\item $\attrel^{\ast}= (\attrel^{C}\cup \attrel')\upharpoonright_{ \argset^{\ast}}$,
\item $(\argset',\attrel')$ is a completion of the underlying cIAF.
\end{itemize}

The notion of control configuration is also straightforwardly adapted to our new class of structures. More in detail, a \textbf{control configuration} of $\ncciaf=\cciaf$ is any $\cfg \subseteq \argset^{\con}$. The CcIAF associated to $\cfg$ is defined as $\ncciaf_{\cfg}=(\cfg,\attrel^{C}\upharpoonright_{ \cfg},\argset^{S}, \varphi)$. {We can extrapolate controllability problems to CcIAFs:}

\begin{center}
\begin{tabular}{|l|}
\hline
$\sigma$-Necessary-Sceptical-Controllability ($\sigma$-NSCon)\\
\hline

\textbf{Given:} A control constrained incomplete argumentation framework  \\
 $\ncciaf=\cciaf$ and an argument $a \in \argset^{S}$ s.th.\ $\models \varphi \to \aw a$. \\

\textbf{Question:} Is it true that there is a configuration \\ $\cfg\subseteq \argset_{\con}$ such that for every completion
$(\argset^{\ast},\attrel^{\ast})$  \\ of $\ncciaf_{\cfg}$ and for every $E\in \sigma(\argset^{\ast},\attrel^{\ast}),a \in E$? \\
\hline
\end{tabular}
\end{center}

Regarding the DL-PA encoding of the reasoning problems that we have just defined, we start by assigning to each CcIAF its \textbf{associated valuation}, in a similar way to what we did both with CAFs and with cIAFs:
\begin{align*}
v_{\ncciaf} & = \attvarset_{\attrel^{C}}
\\&= \{\att x y \mid (x,y)\in \attrel^{C}\}\text{.}
\end{align*}
The control part of a CcIAF is encoded with the same DL-PA program as in CAFs:
\begin{align*}
\control^{\ncciaf}&= \mktruesome(\awset_{\argset^{\con}})\text{.}
\end{align*}
Something analogous happens with the programs for computing completions, where we take over the program we used for cIAFs:
$$\makecomp^{\ncciaf}=\vary(\awset_{\argset^{S}});\vary(\attvarset_{\argset^{S} \times \argset^{S}});\varphi?\text{.}$$

\begin{proposition}
Let $\ncciaf=\cciaf$. 
\begin{itemize}
\item If $(\val_{\ncciaf},\val)\in ||\control^{\ncciaf};\makecomp^{\ncciaf}||$, then there is a control configuration $\cfg\subseteq \argset^{C}$ and a completion $\comp$ of $\ncciaf_{\cfg}$ such that $(\argset_{\val},\attrel_{\val})=\comp$.
\item For every control configuration $\cfg\subseteq \argset^{\con}$ and every $\comp \in \mathsf{completions}(\ncciaf_{\cfg})$ there is a valuation $\val\in 2^{\propset_{\argset^{S}\cup \argset^{\con}}}$  such that $(\val_{\ncaf},\val)\in ||\control^{\ncciaf};\makecomp^{\ncciaf}||$ and $(\argset_{\val},\attrel_{\val})=\comp$.
\end{itemize} \end{proposition}

Once again, we can also reduce reasoning tasks involving CcIAFs to DL-PA model checking problems. Details are left to the reader.

\subsection{Towards a General Theory}\label{sec:generaltheory}
In \cite{DM18}, a general theory of the dynamics of abstract argumentation systems is developed. The focus of the paper is the dynamics of AFs but, as pointed out by the authors, the theory is \emph{prima facie} applicable to other kinds of argumentation frameworks. In this subsection we apply their categorisation to the formalisms for representing qualitative uncertainty about AFs studied in Section~\ref{sec:uncertainty}. At the same time, we show how DL-PA works as a good logical candidate for formalising many parts of this general theory.  

\paragraph{Structural constraints.} According to \cite{DM18}, there are different kinds of constraints that one might want to enforce in an argumentation system. The first kind of constraint is concerned with the structure of an AF. \cite{DM18} distinguishes between \emph{elementary} and \emph{global} structural constraints. 
The former are defined directly on the components of the framework (adding/removing some arguments/attacks); the latter require some property that the output AF must satisfy, e.g., being odd-loop-free, acyclic, etc. Both kinds of constraints make perfect sense in the kind of structures that we have studied in Section~\ref{sec:uncertainty}. Interestingly, the richer nature of these formalisms allows for further distinctions.

\paragraph{Elementary structural constraints.} While in AFs elementary constraints amount to addition/removal of arguments/attacks (or combinations of these, as in the case of AF expansions \cite{BB10}), we can perform more subtle actions in argumentation frameworks with qualitative uncertainty. Let us illustrate some of these actions for the case of IAFs.

\paragraph{Settling uncertain arguments/attacks.}
In a debate, an agent may want to promote the epistemic status of an uncertain argument/attack by ``settling it''. Formally, and restricting our attention to arguments {and incomplete AFs}, given $\niaf=\unraveliaf$ and $a\in \argset^{?}$, define the partial function $\mathsf{settle}:(\mathcal{IAF}\times \uni)\to \mathcal{IAF}$ by:
$$\mathsf{settle}(\niaf,a)=(\argset^{\fix}\cup\{a\},\argset\setminus\{a\}, \attrel^{\fix},\attrel^{?})\text{.}$$
In DL-PA we can compute the completions of the resulting IAF straightforwardly:
$$\compfunc(\mathsf{settle}(\niaf,a))=\{(\argset_v,\attrel_v)\mid (v_{\niaf},v)\in ||\makecomp^{\niaf};\assgntop \aw a||\} . $$

\paragraph{Communicating arguments that become uncertain.} Another kind of dynamics, formally modelled by moving arguments from $\uni \setminus (\argset^{\fix}\cup \argset^{?})$ to $\argset^{?}$, can be used to model situations in which argumentation takes place through a communication channel which is not fully trustworthy (say, a messaging app), so that the proponent is not sure whether the opponent received the arguments that were sent. Again, the completions of the resulting IAF can be easily computed within DL-PA, and the same ideas can be applied to communicating attacks instead of arguments. 

\paragraph{Communicating arguments with uncertain effects.} Yet another kind of action is to communicate arguments whose effects on the opponent's framework are not known. For instance, and within the context of CAFs, one can relax their definition by extending the domain and range of $\attrel^{?}$ or $\attrel^{\leftrightarrow}$ so as to include $\argset^{\con}$.

\paragraph{Belief change methods for logical structures.}
As for cIAFs (and this applies also to cIAFs$^{JM}$), elementary changes amount to either augmenting/shrinking the domain $\argset$ or, more interestingly, changing the epistemic constraint $\varphi$. Regarding the latter, methods imported from the belief change literature can be used; 
for instance, if a new piece of information $\psi$ that is inconsistent with $\varphi$ is to be added, one could do so by means of an AGM belief revision operator \cite{AlchourronEtAl85}. 
In that respect, DL-PA has been shown useful to capture belief change operators \cite{Herzig-Kr14}, and these have been applied in turn to AFs \cite{doutre2014dynamic,doutre2019clar}. \par

\paragraph{Types of elementary structural changes.} Several interesting criteria can be applied to provide a classification of elementary structural changes within frameworks for arguing with qualitative uncertainty. Let us just point out a couple of them. Regarding awareness of arguments, we can distinguish between 
\emph{internal actions}, 
\emph{argument-gaining actions}, and \emph{argument-losing actions}. 
Informally, as the outcome of an internal action, agents neither become aware nor unaware of any new argument.\footnote{However, they might change the epistemic status of arguments they are aware of.} 
A bit more formally, and restricting our attention to IAFs, we say that the action transforming $\niaf_0=(\argset^{F}_0,\argset^{?}_0,\attrel^{F}_0,\attrel^{?}_0)$ into $\niaf_1=(\argset^{F}_1,\argset^{?}_1,\attrel^{F}_1,\attrel^{?}_1)$ is internal whenever $\argset^{F}_0\cup\argset^{?}_0=\argset_1^{F}\cup\argset^{?}_1$. 
For example, the partial function $\mathsf{settle}:(\mathcal{IAF}\times \uni)\to \mathcal{IAF}$ defined above is clearly internal. 
Argument-gaining actions formally amount to requiring that $\argset^{F}_0\cup\argset^{?}_0\subset\argset_1^{F}\cup\argset^{?}_1$, so that the agent becomes aware of at least one novel argument. The action of communicating uncertain arguments that we have explained above is an example of an argument-gaining action. Finally, in argument-losing actions we have that $\argset_1^{F}\cup\argset^{?}_1 \subset \argset^{F}_0\cup\argset^{?}_0$, that is, the agent has become unaware of at least one argument.\footnote{The last type of action connects with a recent thoughtful study of the notion of \emph{forgetting an argument} in the context of AFs \cite{BaumannGR20}. Interestingly, we can capture within IAFs the distinction, made in \cite{forgetting2009}, between \emph{forgetting-as-becoming-unaware} (moving an argument from $\argset^{\fix}\cup \argset^{?}$ to $\uni\setminus(\argset^{\fix}\cup \argset^{?})$), and \emph{forgetting-as-becoming-ignorant} (moving an argument from $\argset^{\fix}$ to $\argset^{?}$).}  A second criterion for categorizing elementary structural changes would be measuring their impact on the number completions, since, intuitively, the more completions we have, the more uncertainty the formalised agent is dealing with. As examples, the $\mathsf{settle}$ function described above always results in a reduction of the number of completions; computing control configurations of CAFs keeps 
the number of completions constant (see Remark~\ref{remark:compcaf} for a precise formulation); and communicating uncertain arguments (also described above) increases the number of completions.

\paragraph{Global structural constraints.} In AFs, global structural constraints amount to things like obtaining an acyclic graph, or an odd-loop-free graph, etc. These constraints are motivated by the appealing mathematical properties implied by them. For instance, it is known since \cite{dung1995acceptability} that in acyclic AFs, all the four classic semantics collapse. Interestingly, DL-PA can capture many of these constraints. For example, in \cite{doutre2019clar} polynomial formulas characterising the existence of odd- and even-length-loops are constructed. 
When extrapolated to the more complex formalisms studied here, global constraints can be required either possibly (that is, in at least one completion) or necessarily (in all of them). Furthermore, DL-PA can be used to check if the constraint is satisfied possibly or necessarily. To be more precise, and focusing on IAFs for simplicity: let $\varphi$ be the formula characterising a targeted global constraint and let $\niaf$ be an IAF; then we have that 
$\varphi$ is satisfied possibly (resp.\ necessary) iff $\val_{\niaf}\models \ldia{\makecomp^{\niaf}}\varphi$ 
(resp.\ iff $\val_{\niaf}\models [\makecomp^{\niaf}]\varphi$). 
In AFs, global constraints are usually enforced through elementary changes (those described above). Once again, this relation can be studied in DL-PA. 
For instance, if we want to know if a global constraint $\varphi$ is possibly enforced in $\niaf$ as the result of settling $a\in \argset^{?}$, it is enough to model-check 
whether $\val_{\niaf}\models \ldia{\makecomp^{\niaf};\assgntop \aw a}\varphi$ holds.
\paragraph{Acceptability constraints.}
The second kind of constraint distinguished by \cite{DM18} is concerned with the output of the argument evaluation process in an argumentation system. The distinction elementary/global applies here, too. When restricted to AFs, one might want to enforce a set of arguments to be part of (or equal to) at least one (or every) extension; this is an elementary acceptability constraint. This kind of enforcement is probably the most studied throughout the literature on abstract argumentation, since the work of \cite{BB10}, as it has a clear informal counterpart in real-life argumentation: persuading an opponent basically amounts to enforcing some targeted arguments. Furthermore, and from a more technical perspective, one might also want to enforce some kind of global acceptability constraint: controlling the cardinality of the set of extensions, its structure, etc. Again, qualitative uncertainty introduces a new layer of quantification: acceptability enforcement can be pursued possibly (i.e., in at least one completion) or necessarily (in all of them). Just as it happens with AFs, acceptability constraints are usually enforced through a (combination of) structural changes such as the ones we have described above. As an example, the reasoning tasks of both CAFs and CcIAFs are a way of enforcing a possible/necessary acceptability constraint through the performance of a combination of elementary structural changes that do not increase uncertainty (activating control arguments). 

\paragraph{Semantic constraints.} Finally, the third kind of constraint distinguished by \cite{DM18} affects the semantics that has been chosen to evaluate arguments. Informally, enforcing a semantic constraint amounts to a change in the standards applied within the argument evaluation process. To this respect, not only the parameter $\sigma$ can be switched to $\sigma'$, but one could also move from credulous to sceptical acceptability, and \emph{vice versa}. Just as before, the formalisms studied in Section~\ref{sec:uncertainty} introduce an additional layer of quantification to be taken into account when formulating semantic constraints: we can move from a `possible' semantics (arguments should be accepted in at least one completion) to a `necessary' semantics (they should be accepted in all), and backwards. As we have shown throughout the paper (e.g., in Proposition~\ref{prop:rediafs}), the distinction between possible and necessary acceptability can be transparently captured in DL-PA.

\section{Discussion, Related Work and Future Directions}\label{sec:final}

We have taken the logical encoding of AFs and their extensions a step further by moving from encodings in propositional logic and \emph{quantified Boolean formulas} (QBF) to encodings in a simple version of dynamic logic DL-PA. 
Approaches to argumentation reasoning problems based on SAT-solvers typically use Besnard and Doutre's encoding of AFs and their semantics in propositional logic \cite{BesnardDoutre}, as well as its extension to QBF for semantics requiring maximality checking; see e.g.\ \cite{DBLP:conf/kr/NiskanenJ20a} for a recent such approach, and \cite{cerutti2017foundations} for a review of approaches to abstract argumentation reasoners.
Based on our work, one could use DL-PA model checkers instead of SAT-solvers in order to automatically decide the reasoning problems that we have investigated here. 
This would however have to await such model checkers, which for the time being do not exist yet. 
Alternatively, one could resort to translations from DL-PA to QBF and use solvers for the latter. 
This is currently pursued in the LILaC group at IRIT.
\par 
On the whole, all we have done in DL-PA can as well be done in equally expressive logical frameworks like propositional logic or QBF.
The advantage over the former is that 
(1)~some semantics can be expressed more compactly in DL-PA, such as the preferred semantics: it is one level higher in the polynomial hierarchy than the other semantics and can therefore not be captured by a polynomial propositional logic formula, while a polynomial DL-PA formula is given in \cite{doutre2019clar},\footnote{Remember that our adaptation of the formula $\mathsf{Preferred}$ of \cite{doutre2019clar} captures preferred semantics in the more general setting of a set of background arguments $\uni$ and is also polynomial in the size of $\uni$.}
and 
(2)~the reasoning problems can be expressed directly as DL-PA programs. 
The advantage over QBFs is that the DL-PA encoding of reasoning problems by means of programs is more natural than the rather complex QBF encodings that one can find in the literature.
Actually, most of the works on arguing with qualitative uncertainty use QBF encodings and algorithms for determining the complexity of associated reasoning tasks (see e.g.\ \cite{baumeister2021acceptance} or \cite{niskanen2021controllability}). All advantages already pointed out by \cite{doutre2019clar} of using DL-PA instead of QBF for encoding argumentative semantics are preserved by our encodings. 
In particular, ``extension construction programs such as $\mathsf{makeExt^{\sigma}}$ capture things in a more general, flexible and natural way than a QBF encoding''. 

\paragraph{Getting closer to a theorem proving approach.} 
Our encoding of formalisms for arguing with qualitative uncertainty can be qualified as \emph{hybrid}, since it combines some previous semantic reasoning with reasoning inside DL-PA. 
For instance, in order to compute the completions of an IAF, one first needs to find its associated valuation (which is reasoning outside the logic, using semantic objects), 
then has to write down the $\mathsf{makeComp}$ program, and finally one reasons in DL-PA to find the $\mathsf{makeComp}$-successors of the associated valuation. 
We followed this hybrid method because we found intuitive the identification of directed graphs with propositional valuations over $\propset$. However, we can adopt results from \cite{doutre2014dynamic,doutre2017dynamic,doutre2019clar} to get a more homogeneous method here. 
For instance, given $\niaf=\iaf$, instead of computing its associated valuation we can write down a propositional formula that characterizes its fixed elements (similar to what is done in \cite{doutre2014dynamic} for standard AFs and in 
our proof of Proposition~\ref{prop:express} in the \nameref{sec:app}): 
$$\mathsf{Th}(\niaf)=\bigwedge_{x \in \argset^{\fix}}\aw x  \land \bigwedge_{x \in \uni \setminus \argset^{\fix}} \lnot \aw x \land \bigwedge_{(x,y) \in \attrel^{\fix}}\att x y \bigwedge_{(x,y) \in \uni \times \uni\setminus \attrel^{\fix}} \lnot \att x y . $$ 
If we combine this formula with the $\mathsf{makeComp}$ program and the converse operator we obtain a formula whose models completely characterize the set of completions of $\niaf$:
\begin{align*}
\mathsf{completions}(\niaf) &= 
\{ (\argset_{v},\attrel_{v})\mid v \in ||\langle \big( \mathsf{Th}(\niaf)?;\mathsf{makeComp}^{\niaf}\big)^{\smallsmile}\rangle \top||\} .
\end{align*}

\paragraph{Novel contents w.r.t.\ our conference paper \cite{clar2021}.} 
This work is based on our previous conference paper \cite{clar2021}, which we have improved and extended in three main different directions. First, in Section~\ref{sec:semantics}, 
(i) we capture argumentation semantics in DL-PA that had not been captured before (naive, semi-stable, stage, ideal and eager semantics), and (ii) we also adapt previous encodings to our more general setting (in particular, we adapt the encodings of complete, preferred and grounded semantics \cite{doutre2019clar} to the assumption of the existence of a background universe of arguments $\uni$, which is in turn useful for modelling both dynamics and uncertainty about AFs). Second, we discussed some closely related works that appeared since (sections~\ref{sec:related} and \ref{sec:comparison}). Third, we provided new results regarding the combination of dynamics and uncertainty in abstract argumentation: sections~\ref{sec:cciafs} and \ref{sec:generaltheory} are entirely new. Finally, there are also several small improvements w.r.t.\ the conference version, some of which are signalled throughout the paper.
\paragraph{Epistemic aspects of argumentation.}
In recent years, a few papers dealing with the combination of epistemic logic and formal argumentation have appeared. Broadly speaking, these works can be divided into two main branches: 
(i)~those trying to provide a formalisation of the notion of justified belief based on argumentative tools such as \cite{grossi2014,shi2017argument,shi2018,shi2021logic,comma,tark}; and 
(ii)~those using epistemic models for reasoning about uncertain AFs such as \cite{schwarzentruber2012building, proietti2021, ijcai21, kr}. 
Clearly, the second one is strongly connected---both conceptually and technically---to some of the ideas presented here. The main first difference is that the formalisms used in this paper lack a tool for capturing higher-order epistemic attitudes, that is, a tool capable of representing not only what an agent thinks of her opponent's argumentative situation (her AF), but also about what the agent thinks that her opponent thinks about the agent's argumentative situation, and so on. This is an important point, since this kind of mental attitude has been successfully employed under the name of \emph{recursive opponent models} within the sub-field of strategic argumentation (see, e.g., \cite{thimm2014strategic}). However, the incorporation of this type of multiple agency together with a full dynamic tool-kit would mean to replace DL-PA by the strong modelling power of dynamic epistemic logic \cite{DitmarschHoekKooi07}. 
This comes at the price of a blow-up in the computational complexity of the associated reasoning tasks. 
One might however follow \cite{DBLP:journals/ai/CooperHMMPR21} and employ lightweight epistemic logics where disjunctions in the scope of epistemic operators are forbidden.
That would represent a compromise between modelling multiple agency/dynamics, on the one side, and modelling uncertainty, on the other side, since any form of uncertainty that goes beyond IAFs (see Figure \ref{fig:exp}) would have to be excluded from this approach. {A second important difference is that, unlike epistemic logic, none of the formalisms studied in this paper allow for modelling \emph{the actual world}, i.e., what is true independently of what the formalised agent thinks. This notion is in turn needed for distinguishing between \emph{knowledge} (which is usually required to be true) and \emph{belief} (which is often merely required to be consistent). However, this limitation seems easier to be overcome: It suffices to augment IAFs (and their extensions) with a \emph{distinguished completion}, informally accounting for what the actual AF is.}
\paragraph{Further semantics.} Yet another direction for future work is extending our DL-PA encoding to semantics that have not been considered in Section~\ref{sec:semantics}. A specially interesting case is the recently introduced family of weak admissibility-based semantics \cite{BaumannBU20}, since most of the associated reasoning tasks have been shown to be PSPACE-complete, matching the complexity of the DL-PA model checking problem \cite{BalbianiHST14}.
\paragraph{An alternative notion of expressivity.} In a very recent paper \cite{alfano2022aaai}, Alfano et al.\ invented a rewriting technique in order to reduce general IAFs to their strict subclasses arg-IAFs and att-IAF, and yet to a proper subclass of arg-IAFs.\footnote{The so-called \emph{fact-uncertain AFs} (farg-IAFs), which are argument-incomplete AFs where all uncertain arguments are not attacked.} More concretely, they show \cite[Theorem 7]{alfano2022aaai} that the completions of the rewritten incomplete AF can be mapped (through another transformation) to the completions of the original one. They moreover claim that ``This result entails that arg-IAFs (resp.\ farg-IAF, att-IAF) have the same expressivity of general IAFs, though arg-IAFs (resp.\ farg-IAF, att-IAF) have a simpler structure''. This clearly conflicts with the expressivity map that we provided in Proposition \ref{prop:ex} and Figure \ref{fig:exp}, which is based on the notion of expressivity first introduced in \cite{mailly2020note} and later used in \cite{mailly2021yes,clar2021}. Although a detailed comparison of both notions of expressivity is out of the scope of this discussion, we would just like to mention that the one employed here seems more useful for intuitive modelling purposes (i.e., to find out what kind of situations the formalised agent is able to represent in her mind), while Alfano et.\ al's seems more interesting from a technical perspective (actually, it is used to extend complexity results regarding IAFs to their proper subclasses). Be as it may, the work done in \cite{alfano2022aaai} opens an interesting research question: can the rewriting technique be extended to more expressive formalisms (in our sense), such as rIAFs or cIAFs?

\black

\subsubsection*{Funding}
The research activity of both authors is partially supported by the EU ICT-48 2020 project TAILOR (No. 952215). Part of this research was carried on when Antonio Yuste was employed by the University of M\'alaga through a Post-doctoral contract type A.3.1.\ of \emph{Plan Propio de Investigaci\'on, Transferencia y Divulgaci\'on Cient\'ifica}.

\subsubsection*{Acknowledgements}
We thank Sylvie Doutre and Jean-Guy Mailly for previous discussions on the topic of this paper, specially for triggering the idea of constrained incomplete argumentation frameworks.

\bibliographystyle{plain}
\bibliography{argpi_biblio}

\section*{Appendix}\label{sec:app}
In this appendix, we provide selected proofs and proof sketches for the results found throughout the paper. \par \medskip

\noindent[\textbf{Theorem \ref{thrm:encodings}}]

\begin{proof}

We shall just prove the first bullet for $\sigma=se$. The other cases are simpler and follow similar arguments, while the second bullet follows easily from the first one and the meaning of $\mkext$.
Let us first state without proof a couple of needed lemmas

\begin{lemma}\label{lemma:copy}
$(\val,\val_1) \in ||\mathsf{copy}(\inset_{\uni});\mkext^{\sigma}||$ implies $\extOf\val=\{x \mid \acc x ' \in \val_1\}$ and $(\argset_\val,\attrel_\val)=(\argset_{\val_1},\attrel_{\val_1})$.
\end{lemma}

\begin{lemma}\label{lemma:copyinclu} 
Let $\val \subseteq \propset$ s.th.\ $\val\models \well$ and $\{x \mid \acc x '\in \val\}\subseteq \{x \mid \aw x \in \val\}$, then:
\begin{itemize}
\item $\val \models \mathsf{IncludesCp}$ iff $\{x \mid \acc x ' \in \val\}^{\oplus}\subseteq \extOf \val^{\oplus}$.
\item $\val \models \mathsf{IncludedInCp}$ iff $ \extOf \val^{\oplus} \subseteq \{x \mid \acc x ' \in \val\}^{\oplus}$.
\end{itemize}

\end{lemma}

\noindent ($\Rightarrow$) Suppose that $\val \models \mathsf{Semistable}$, which amounts to\\
$\val \models \complete$ and $\val \models \lbox{ \mathsf{copy}(\inset_{\uni}) ; \mkext^{co} } \left(
 \mathsf{IncludesCp} \limp \mathsf{IncludedInCp}  \right)$. The first conjunct is equivalent, by the same item we are proving but for $\sigma=co$, to $\extOf \val \in \mathsf{co}(\argset_v,\attrel_v) $. So, we just need to show that $\extOf \val$ has a maximal range among complete extensions. Suppose $E'\in \mathsf{co}(\argset_\val,\attrel_v)$. By by the same item that we are proving but for $\sigma=co$ and Lemma~\ref{lemma:copy}, we have that $E'=\extOf{\val_1}$ and $\extOf\val=\{x \mid \acc x ' \in \val_1\}$ for some $(\val,\val_1)\in ||\mathsf{copy}(\inset_{\uni});\mkext^{co}||$. Suppose that $\extOf\val^{\oplus}\subseteq \extOf{\val_1}^{\oplus}$. Note that $\val_1$ satisfies the antecedent of Lemma~\ref{lemma:copyinclu} (this is deducible from $\val\models \complete $ and Lemma~\ref{lemma:copy}). Hence, we have that $\extOf\val^{\oplus}\subseteq \extOf{\val_1}^{\oplus}$ is equivalent to $\val_1\models \mathsf{IncludesCp}$. Since we know that $\val_1\models \mathsf{IncludesCp} \to \mathsf{IncludedInCp}$, we can deduce $\val_1 \models \mathsf{IncludedInCp}$, which by Lemma~\ref{lemma:copyinclu} again, amounts to $\extOf{\val_1}^{\oplus}\subseteq \extOf{\val}^{\oplus}$. Since $\extOf{\val_1}$ was an arbitrary complete extension of $(\argset_v,\attrel_v)$, we can conclude that the range of $\extOf \val$ is maximal among the ranges of complete extensions.
 
 \par
 \newcommand{\afOf}[1]{(\argset_{#1},\attrel_{#1})}
 ($\Leftarrow$) Suppose that $\extOf\val \in \mathsf{se}(\afOf\val)$, which amounts to\\
 (i) $\extOf\val \in \mathsf{co}\afOf\val$ and (ii) the range of $\extOf\val$ is maximal among those of the complete extensions of $\afOf\val$. From (i) and the same item we are proving but for $\sigma=co$, we obtain $\val\models \complete$. Hence we just need to show that the second conjunct of $\mathsf{Semistable}$ is true at $\val$. For doing so, suppose that $(\val,\val_1)\in ||\mathsf{copy}(\inset_{\uni});\mkext^{co}||$ and $\val_1\models \mathsf{IncludesCp}$. From both lemmas and the previous assertion, we can arrive to $\val_1 \models \mathsf{IncludedInCp}$.
 
\end{proof}
\par \medskip
\noindent [\textbf{Proposition \ref{prop:iafenco}}]
\begin{proof}
For the first item, suppose $(\val_{\niaf},\val)\in ||\makecomp^{\niaf}||$. 
We recall from Proposition \ref{prop:dlpaprg} that $||\mktruesome(\mathsf{P})||=\{(\val',\val'')\mid \val''=\val'\cup S, S\subseteq \mathsf{P}\}$ for any set of atoms $\mathsf{P}$.
By the semantics of the sequential composition operator ``$;$'', $(\val_{\niaf},\val)\in ||\makecomp^{\niaf}||$ amounts to saying
that $\val=\val_{\niaf}\cup \mathsf{P}$ for some $\mathsf{P}\subseteq \awset_{\argset^{?}}\cup \attvarset_{\attrel^{?}}$. From this statement, and applying the definition of $(\argset_v,\attrel_v)$ and the one of completion, we obtain 
that $(\argset_{\val},\attrel_{\val})\in \mathsf{completions}(\niaf)$. 
 
For the second item, suppose that $(\argset^{\ast},\attrel^{\ast})\in \mathsf{completions}(\niaf)$, which amounts to $\argset^{\fix} \subseteq \argset^{\ast}\subseteq \argset^{\fix}\cup \argset^{?}$ and $\attrel^{\fix}\upharpoonright_{ \argset^{*}}\subseteq \attrel^{\ast}\subseteq (\attrel^{\fix}\cup\attrel^{?})\upharpoonright_{ \argset^{\ast}}$. Now, remember that $\val_{\comp}=\awset_{\argset^{\ast}}\cup \attvarset_{\attrel^{\ast}}$. From the two previous statements and the definition of $\val_{\niaf}$, we can deduce that the set of variables whose truth values differ from $\val$ to $\val_{\comp}$ must be a subset of $\awset_{\argset^{?}}\cup \attvarset_{\attrel^{?}}$, which, as argued before, amounts to saying that $(\val_{\niaf},\val_{\comp})\in ||\makecomp^{\niaf}||$.
\end{proof}

\par \medskip

\noindent[\textbf{Proposition \ref{prop:rediafs}}]

\begin{proof}[Sketch of proof] The result follows from the definition of the reasoning task, the correctness of each $\mkext^{\sigma}$ (Theorem~\ref{thrm:encodings}), Proposition~\ref{prop:iafenco}, and the interpretation of $\lbox{.}$ and $\ldia{.}$ in DL-PA.
\end{proof}

\noindent[\textbf{Proposition \ref{prop:riafenco}}]
\begin{proof}[Sketch of proof] The proof is analogous to that of Proposition~\ref{prop:iafenco}, but takes into account the observation that, when applied to the \emph{symmetric} relation $\attrel^{\leftrightarrow}=\{(x_1,y_1), (y_1,x_1),...,(x_n,y_n),(y_n,x_n)\}$, every execution of $\dis(\attvarset_{\attrel^{\leftrightarrow}})$ makes true either $\att {x_i} {y_i}$, or $\att {y_i} {x_i} $ or both, for every $1\leq i \leq n$. This ensures that the last clause of the definition of completion for rIAFs is captured in the DL-PA program $\makecomp^{\nriaf}$.
\end{proof}

\par \medskip
\noindent [\textbf{Proposition \ref{prop:redriafs}}]
\begin{proof}[Sketch of proof]
The result follows from the definition of the reasoning problem, the correctness of $\mkext^{\sigma}$ (Theorem~\ref{thrm:encodings}), the correctness of $\makecomp^{\nriaf}$ (Proposition~\ref{prop:riafenco}), and the semantics of DL-PA.
\end{proof}

\noindent [\textbf{Proposition \ref{prop:encociafs}}]
\begin{proof}[Sketch of proof]  
The interpretation of $\vary(\awset_{\argset});\vary(\attvarset_{\argset \times \argset})$, when restricted to $2^{\propset_{\argset} \setminus  \inset_{\argset}}$, is actually the total relation $2^{\propset_{\argset} \setminus  \inset_{\argset}} \times 2^{\propset_{\argset} \setminus  \inset_{\argset}}$. 
Hence from $ \val_{\nciaf} = \emptyset$
we have an execution of $\vary(\awset_{\argset});\vary(\attvarset_{\argset \times \argset})$ that goes to \emph{any} valuation in $2^{\propset_{\argset} \setminus  \inset_{\argset}}$. Then the execution of $\varphi?$
filters those valuations of $2^{\propset_{\argset} \setminus  \inset_{\argset}}$ that satisfy the constraint of $\nciaf$, i.e., the set of valuations of $2^{\propset_{\argset} \setminus  \inset_{\argset}}$ representing the set of completions of $\nciaf$. 

\end{proof}
\par \medskip
\noindent [\textbf{Proposition \ref{prop:redciaf}}]
\begin{proof}[Sketch of proof] The result follows from from the definition of the reasoning task, the correctness of $\mkext^{\sigma}$ (Theorem~\ref{thrm:encodings}), Proposition~\ref{prop:encociafs}, and the semantics of DL-PA. 
\end{proof}

\par \medskip

\noindent[\textbf{Proposition \ref{prop:express}}]

\begin{proof} 
We only have to prove $ c\text{-}\mathcal{IAF} \succ \mathcal{RIAF}$ because $c\text{-}\mathcal{IAF} \succ \mathcal{IAF}$ follows from $\mathcal{RIAF} \succ \mathcal{IAF}$ \cite{mailly2020note} and the transitivity of $\succ$.

To prove $ c\text{-}\mathcal{IAF} \succeq \mathcal{RIAF}$, suppose $\nriaf$ is a rIAF with $\mathsf{completions}(\nriaf)=\{(\argset^{\ast}_1,\attrel^{\ast}_1),...,(\argset^{\ast}_n,\attrel^{\ast}_n)\}$. 
For every AF $(\argset,\attrel)$ defined over $\uni$ we can write its \emph{theory} (see e.g.\ \cite{doutre2017dynamic}), that is, the propositional formula
$$\mathsf{Th}(\argset,\attrel)=\bigwedge_{x \in \argset}\aw x \land \bigwedge_{x \in \uni \setminus \argset} \lnot \aw x \land \bigwedge_{(x,y) \in \attrel} \att x y \land \bigwedge_{(x,y) \in \uni \times \uni\setminus \attrel} \lnot \att x y \text{.} $$
It is then easy to show that for any valuation $\val \subseteq \propset$, we have that $\val \models \mathsf{Th}(\argset,\attrel)$ iff $(\argset_{\val},\attrel_{v})=(\argset,\attrel)$.  Now, letting $\rho=\bigvee_{1\leq i \leq n}\mathsf{Th}(\argset^{\ast}_{i},\attrel^{\ast}_{i})$, we have that

$$\mathsf{completions}(\uni,\rho)=\mathsf{completions}(\nriaf)\text{.}$$

In order to prove 
that $ \mathcal{RIAF} \not\succeq c\text{-}\mathcal{IAF}$
it suffices to show that the cIAF of Example~\ref{ex:ciaf} (called $\nciaf_0$) cannot be expressed as a rIAF. 
Reasoning towards a contradiction, suppose that there is a rIAF $\nriaf=\riaf$ with the same set of completions as $\nciaf_0$. Then we would have $(a,b)\in \attrel^{\fix}\cup\attrel^{?}\cup \attrel^{\leftrightarrow}$ (since $(a,b)$ appears in a completion of $\nriaf$). We show that the last statement is absurd. 
If $(a,b)\in \attrel^{\fix}$ then $(a,b)$ should appear in all completions of $\nriaf$ where $a$ and $b$ are present, but this is not true. 
If $(a,b) \in \attrel^{?}$ then we reason by cases on $(b,a)\in \attrel^{\fix}\cup\attrel^{?}\cup \attrel^{\leftrightarrow}$: the first one is impossible, since $(b,a)$ would be in every completion where $a$ and $b$ appear, and that is not the case; the second one is absurd because we would have an extension with neither $(a,b)$ nor $(b,a)$, and this is not the case; the third one is impossible because it would imply $(a,b)\in \attrel^{\leftrightarrow}$, but we have assumed that $(a,b)\in \attrel^{?}$, and we know that $\attrel^{?}\cap \attrel^{\leftrightarrow}=\emptyset$ by definition. Finally, suppose that $(a,b)\in \attrel^{\leftrightarrow}$, which implies $(b,a)\in \attrel^{\leftrightarrow}$ (by symmetry of $\attrel^{\leftrightarrow}$), which is impossible because we would have a completion containing both $(a,b)$ and $(b,a)$, but this is not the case.
\end{proof}
\par \medskip
\noindent[\textbf{Proposition \ref{prop:cafsenco}}]
\begin{proof}[Sketch of proof] The proof is analogous to those of propositions~\ref{prop:iafenco} and \ref{prop:riafenco}. The essential difference lies in the fact that the previous execution of $\control^{\ncaf}$ is needed to nondeterministically choose a control configuration of $\ncaf$. Also, note that $\attvarset_{\attrel^{\con}}\subseteq \val_{\ncaf}$ is essential for obtaining the needed control attacks in the corresponding completion.
\end{proof}

\noindent[\textbf{Proposition \ref{prop:redcaf}}]
\begin{proof}[Sketch of proof]
The result follows from the definition of the reasoning task, the correctness of $\mkext^{\sigma}$ (Theorem~\ref{thrm:encodings}), Proposition~\ref{prop:cafsenco}, and the semantics of DL-PA.
\end{proof}
\end{document}